\documentclass[12pt]{article}

\usepackage[margin=1in]{geometry}
\usepackage{amsmath,amssymb,mathrsfs}
\usepackage{booktabs}
\usepackage[sort]{cite}
\usepackage{hyperref}
\usepackage{graphicx}
\usepackage[section]{placeins}
\usepackage{xcolor}
\usepackage{parskip}
\usepackage{enumitem}
\usepackage{aas_macros}
\usepackage[affil-it]{authblk}
\let\affiliation\affil

\hypersetup{colorlinks=true,linkcolor=blue,urlcolor=blue,citecolor=blue}

\newcommand{\rmin}{r_{\mathrm{m}}}

\newcommand{\rin}{r_{\mathrm{in}}}
\newcommand{\rout}{r_{\mathrm{out}}}
\newcommand{\bcrit}{b_{\mathrm{c}}}
\newcommand{\beff}{b_{\mathrm{eff}}}

\newcommand{\dd}{\mathrm{d}}

\begin{document}

\title{The Glow of Eternal Black Holes}

\author[1,2,$\ast$]{Vladimir Strokov} %
\affiliation[1]{Department of Physics and Astronomy, West Virginia University, USA}
\affiliation[2]{Center for Gravitational Waves and Cosmology (GWAC), West Virginia University, USA}

\date{\today}

\maketitle
\begingroup
\renewcommand{\thefootnote}{$\ast$}
\footnotetext{\href{mailto:vladimir.strokov@mail.wvu.edu}{vladimir.strokov@mail.wvu.edu}}
\endgroup

\begin{abstract}
We compute the optical appearance of a maximally extended (eternal) Schwarzschild black hole surrounded by a geometrically thin accretion disk. Unlike astrophysical black holes formed by gravitational collapse, the eternal solution contains a white hole (WH) region connected to a past singularity. Following Markov's hypothesis of a limiting density of matter, we replace the past singularity with a spacelike ``surface of last scattering'' at $r=\rmin$ and assume that it emits black-body radiation with temperature~$T$. Past-directed rays that cross the past horizon terminate on this surface, resulting in a bright spot at the center of the shadow of an eternal black hole. The radial profile of the spot is set by the gravitational frequency shift, with a blueshifted center when the spacelike surface is sufficiently close to the singularity. We use ray tracing to generate an image of the disk and the spot, and we derive a closed-form interferometric signature of the spot. Its visibility is a pure exponential in the baseline length, in contrast to the power-law envelopes of the disk and the photon ring. We constrain the product $\rmin T$ from the observed 230\,GHz fluxes of M87$^\star$ and Sgr~A$^\star$. We also consider a Planck-star scenario in which the value of $\rmin$ is determined on dimensional grounds, and show that it is out of reach of any current or planned facilities. Nevertheless, the darkness of the observed shadows remains a test of whether these black holes are eternal.
\end{abstract}

\section{Introduction}
\label{sec:intro}

The idea of a body so compact that light cannot escape its surface dates back to Michell~\cite{Michell:1784xqa} and Laplace~\cite{Laplace:1796}. But it was not until Schwarzschild found a spherically symmetric solution to Einstein's equations~\cite{Schwarzschild:1916uq} that the event horizon of a black hole of mass~$M$ was formally defined at~$r_{\rm h} = 2GM/c^2$ ($G$ is the gravitational constant and $c$ is the speed of light). Because a classical black hole emits no light of its own, its appearance must be inferred from its silhouette against a luminous background. Light rays that fall inside the horizon leave a deficit on the observer's sky, known as the black hole ``shadow''. Its size and shape are dictated entirely by the spacetime geometry, in particular by the unstable photon orbits that surround the horizon~\cite{Misner:1973prb,chandrasekhar1983,Perlick:2021aok,Cunha:2018acu}. Another geometric feature is a narrow ``photon ring'' formed by light rays that complete one or more half-orbits before escaping to infinity. Already the first ray-traced images of thin accretion disks resolved this ring into a sequence of exponentially demagnified subrings ($n = 1, 2, \ldots$)~\cite{Luminet:1979nyg,Luminet:2019hfx}. Since the subring shape is determined entirely by the black hole mass and spin, it is essentially independent of the details of the accretion flow~\cite{Gralla:2019xty,Johannsen:2010ru,Gralla:2020yvo,Himwich:2020msm}. By contrast, the broader image structure depends on the physical properties of the accreting plasma and varies substantially across semi-analytic and general relativistic magnetohydrodynamic (GRMHD) emission models~\cite{Bromley:2001er,Broderick:2005jj,Noble:2007zx,Broderick:2008qf,Moscibrodzka:2009gw,Dexter:2011xa,Dibi:2012tq,Lu:2014zja,Chan:2015kpa,Moscibrodzka:2015pda,Porth:2016rfi,Chael:2018aeq,Ryan:2018zhq,Davelaar:2019jxr}.

Even for the largest supermassive black holes, resolving the horizon scale requires very-long-baseline interferometry (VLBI) at submillimeter wavelengths~\cite{Falcke:1999pj}. At these frequencies, the radiatively inefficient accretion flows around Sgr~A$^\star$ and M87$^\star$ become optically thin, revealing the horizon scale~\cite{Ozel:2000wm,Broderick:2008qf}. A multi-year effort by a global network of radio observatories known as the Event Horizon Telescope (EHT)~\cite{Doeleman:2008qh,Doeleman:2012zc,Akiyama:2015qta} culminated in the first resolved images of M87$^\star$ and Sgr~A$^\star$~\cite{EventHorizonTelescope:2019dse,EventHorizonTelescope:2022wkp,EventHorizonTelescope:2022wok}, thus providing direct insight into strong-field gravity. Subsequent polarimetric follow-ups~\cite{EventHorizonTelescope:2021bee,EventHorizonTelescope:2021srq,EventHorizonTelescope:2024hpu} have mapped their horizon-scale magnetic-field structures. And multi-epoch analyses confirm that the M87$^\star$ ring diameter remains stable across several years~\cite{EventHorizonTelescope:2025vum}, though its measured ellipticity is dominated by turbulent accretion rather than spacetime geometry~\cite{EventHorizonTelescope:2025whi}.

Proposals for the next generation of horizon-scale interferometry span a few directions, such as imaging other supermassive black holes~\cite{Faggert:2025eja}, improving the angular resolution of the existing M87$^\star$ and Sgr~A$^\star$ images, and detecting the photon ring directly. The next-generation EHT (ngEHT)~\cite{Doeleman:2023kzg,Ayzenberg:2023hfw} aims to expand the global array on the ground and to operate at higher frequencies. The Black Hole Explorer (BHEX) mission~\cite{Johnson:2024ttr,Lupsasca:2024xhq} is proposed as a dedicated probe of the photon-ring shape and involves space VLBI. Baselines exceeding the Earth diameter are crucial, as the photon ring is predicted to have a distinctive signature in its visibility amplitude at long baselines~\cite{Johnson:2019ljv}. This signature is robust across Kerr parameters~\cite{Gralla:2020srx,Cardenas-Avendano:2023dzo,Paugnat:2022qzy}, finite-width corrections~\cite{Jia:2024mlb}, and instrument noise and plasma fluctuations~\cite{Cardenas-Avendano:2024flu} (with potential false positives from hybrid image reconstruction~\cite{Tiede:2022grp}). Complementary space--space VLBI concepts use multiple orbiting stations to reach the higher resolution~\cite{Roelofs:2019nmh,Zineb:2024gwx} needed to image a broader sample of nearby supermassive black holes~\cite{Pesce:2021adg,Zhang:2024owe}. Rapid $uv$-plane coverage also opens the possibility of dynamical imaging of variable sources such as Sgr~A$^\star$~\cite{2019ApJ...881...62P,Zineb:2024gwx}.

The 2022 EHT image of Sgr~A$^\star$ also provides a test of general relativity, which translates the measured ring diameter into limits on regular black holes, no-hair violations, modified-gravity solutions, and horizonless mimickers including wormholes and naked singularities~\cite{EventHorizonTelescope:2022xqj,Vagnozzi:2022moj,Carballo-Rubio:2022imz}. Similar bounds can be derived from the 2019 M87$^\star$ image~\cite{EventHorizonTelescope:2020qrl,EventHorizonTelescope:2021dqv}. More generally, shadow morphology serves as a probe of the spacetime and a direct tool for constraining deviations from the Kerr metric~\cite{Perlick:2021aok,Cunha:2018acu}. A number of ray-traced shadows have been computed for alternative compact objects, including rotating black holes in Chern--Simons gravity~\cite{Amarilla:2010zq}, traversable wormholes~\cite{Nandi:2006ds,Bambi:2013nla,Nedkova:2013msa,Ohgami:2015nra,Shaikh:2018kfv,Bugaev:2021dna}, regular black holes with curvature-bounded interiors~\cite{Stuchlik:2019uvf,Kumar:2020yem,Eichhorn:2021iwq,Eichhorn:2022oma}, and horizonless ultracompact objects such as boson and Proca stars~\cite{Vincent:2015xta,Olivares:2018abq,Herdeiro:2021lwl}. \textit{Notably, the shadow of a compact object is not necessarily dark}. For example, in the case of traversable wormholes, the throat transmits light from a second asymptotic region, producing a bright interior image whose morphology directly reflects the throat geometry (e.g.,~\cite{Shatskiy:2008ym}).

In this work, we consider a similar yet distinct possibility. The maximally extended (eternal) Schwarzschild solution contains a white hole (WH) region connected to a past singularity. Although ordinary stellar gravitational collapse cannot produce this geometry, objects similar to the eternal Schwarzschild black hole could be artifacts formed in the early universe. Examples of such objects are static regular black-hole metrics with curvature-bounded interiors~\cite{Frolov:2016pav,DeLorenzo:2014pta}, black-to-white-hole tunneling scenarios in which matter re-emerges across the past horizon~\cite{Lukash:2013ts,Rovelli:2014cta,Haggard:2014rza,Barrau:2014yka,DeLorenzo:2015gtx,Bianchi:2018mml,Rovelli:2018okm,BenAchour:2020gon}, and loop quantum gravity quantizations of the {Kruskal} interior~\cite{Corichi:2015xia,Ashtekar:2018lag,Ashtekar:2020ckv}. For the present analysis, we take this geometry at face value. Since the presence of singularities implies a breakdown of classical general relativity, \textit{we swap the past singularity for a region bounded by an emitting spacelike surface}. This is in the spirit of Markov's hypothesis of a limiting density of matter as a ``universal law of nature''~\cite{Markov:1982,Markov:1984ii}. The size of the region can be estimated by requiring a Planckian value for the Kretschmann scalar~\cite{Frolov:1998wf}, though we keep it arbitrary in our calculations.

Here we argue that this assumption produces a distinctive optical and interferometric signature. Past-directed rays crossing the horizon terminate on the emitting surface, resulting in a bright spot at the center of the shadow. To the best of our knowledge, this signature has not been previously studied. Using the Luminet ray-tracing method~\cite{Luminet:1979nyg}, we compute the observer-sky image and the interferometric visibility of the WH spot for a thin accretion disk. If we also assume that the radiation from the surface follows the Planck law, the gravitational frequency shift (both blue- and redshift) sets the spot's surface brightness. As it turns out, this leads to a simple analytical closed form for the visibility of the spot: a pure exponential in the baseline length. We also compare the visibility profiles of the bright spot with those of the direct disk image and the $n = 1$ photon ring.

The paper is organized as follows. In Section~\ref{sec:spacetime} we review the Penrose diagrams of the eternal and astrophysical Schwarzschild spacetimes but present them in a form that is more suitable for illustrating the ray-tracing. Section~\ref{sec:brightness} computes the WH bright spot and the observer-sky image of the eternal geometry with a thin accretion disk. In Section~\ref{sec:visibility} we present the interferometric signatures of the direct disk image, the $n = 1$ photon ring, and the WH spot. For the spot, we derive a closed-form analytical expression for the visibility and discuss its physical normalization at 230\,GHz. Section~\ref{sec:discussion} summarizes the results and discusses limitations, observational degeneracies, and future prospects. If not stated otherwise, we use geometric units $G=c=1$.

\section{Observer-friendly Penrose diagrams}
\label{sec:spacetime}

In this section we review the Penrose diagrams of the eternal and astrophysical Schwarzschild spacetimes and use them to display the null geodesics on which the ray-tracing of the subsequent sections is based (see Section~\ref{sec:image}). Here we view the diagrams from a perspective that is somewhat different from what is usually adopted in the literature. First, we use a scale factor in the compactification of the Kruskal--Szekeres coordinates in order to ``zoom in'' on larger radii, i.e., the radii that are important for ray-traced images as seen by a distant observer. And second, we show not only radial null geodesics but also the projections of light rays with nonzero angular momenta onto the plane of the diagram. To give the reader some intuition about the appearance of such rays, let us start from the vacuum Schwarzschild metric ($M=1$ is assumed throughout the paper)
\begin{equation}\label{eq:schw_metric}
    ds^2 = \left(1 - \frac{2}{r}\right) \dd t^2 - \left(1 - \frac{2}{r}\right)^{-1} \dd r^2 - r^2\, \dd\Omega^2,
\end{equation}
where $\dd\Omega^2 = \dd\theta^2 + \sin^2\theta\, \dd\varphi^2$, and rearrange the null condition $ds^2 = 0$ into
\begin{equation}\label{eq:null_projection}
    \left(1 - \frac{2}{r}\right) \dd t^2 - \left(1 - \frac{2}{r}\right)^{-1} \dd r^2 = r^2\, \dd\Omega^2 \geq 0.
\end{equation}
Since the right-hand side is non-negative, non-radial null geodesics \emph{appear} timelike (steeper than $45^\circ$) in the diagrams that follow. Of course, purely radial rays ($\dd\Omega = 0$) still travel along the light cones of the diagram. 

Below we present a progression of Penrose diagrams, from the standard textbook compactification to a globally rescaled, observer-adapted version. Together they make the case that, for an observer far in the future of a gravitational collapse, the past geodesic histories in the eternal and astrophysical spacetimes are virtually indistinguishable. However, the qualitative difference remains: in the eternal case, rays emerging from the past horizon originate at the WH singularity.

Penrose diagrams conformally compactify an asymptotically flat spacetime to a finite region of the plane while preserving its causal structure~\cite{Misner:1973prb,chandrasekhar1983}: radial null rays travel at $45^\circ$, and both null infinity $\mathscr{I}^\pm$ and spatial infinity $i^0$ are mapped to finite boundaries. The $(t, r)$ chart of Eq.~\eqref{eq:schw_metric} breaks down at the event horizon $r = 2$. In order to extend the geometry smoothly through it, one introduces Kruskal--Szekeres coordinates $(U, V)$, related to $(t, r)$ in the right exterior ($r > 2$) by
\begin{equation}\label{eq:ks_from_schw}
    U = \sqrt{\frac{r}{2} - 1}\; e^{r/4} \cosh(t/4), \qquad
    V = \sqrt{\frac{r}{2} - 1}\; e^{r/4} \sinh(t/4),
\end{equation}
with inverse
\begin{equation}\label{eq:schw_from_ks}
    r = 2\!\left[1 + W_0\!\left(\frac{U^2 - V^2}{e}\right)\right], \qquad
    t = 2\,\ln\left|\frac{V + U}{V - U}\right|,
\end{equation}
where $W_0$ denotes the principal branch of the Lambert $W$ function. Analytic continuation of Eqs.~\eqref{eq:ks_from_schw}--\eqref{eq:schw_from_ks} across the horizons uncovers four causally distinct regions: the right exterior (Region~I) where the observer resides, the black hole interior (Region~II) containing the future spacelike singularity $r = 0$, a second asymptotic exterior (Region~III) causally disconnected from Region~I, and the WH interior (Region~IV) containing a spacelike singularity in the past. In the $(U, V)$ chart the metric reads
\begin{equation}\label{eq:ks_metric}
    ds^2 = \frac{32}{r}\, e^{-r/2}\, (\dd V^2 - \dd U^2) - r^2\, \dd\Omega^2,
\end{equation}
manifestly regular at $r = 2$, with $V$ timelike. A one-parameter family of compactifications maps the infinite $(U, V)$ plane to the finite diamond $\chi, \eta \in [-\pi/2,\,\pi/2]$:
\begin{align}
    \eta &= \frac{1}{2}\left[\arctan\!\left(\frac{V+U}{S}\right) + \arctan\!\left(\frac{V-U}{S}\right)\right], \label{eq:compact_eta}\\
    \chi &= \frac{1}{2}\left[\arctan\!\left(\frac{V+U}{S}\right) - \arctan\!\left(\frac{V-U}{S}\right)\right], \label{eq:compact_chi}
\end{align}
with inverse
\begin{equation}\label{eq:compact_inverse}
    V \pm U = S\,\tan(\eta \pm \chi),
\end{equation}
where $S$ is a global scale. The standard textbook diagram corresponds to $S = 1$, and below we use $S > 1$ for an observer-centric view of the diagrams. For any $S$, null infinity sits on the diagonal edges $\eta \pm \chi = \pm\pi/2$, and the event horizons at $r = 2$ run as $45^\circ$ null lines through the center.

\begin{figure}[htbp]
\centering
\includegraphics[width=0.8\textwidth]{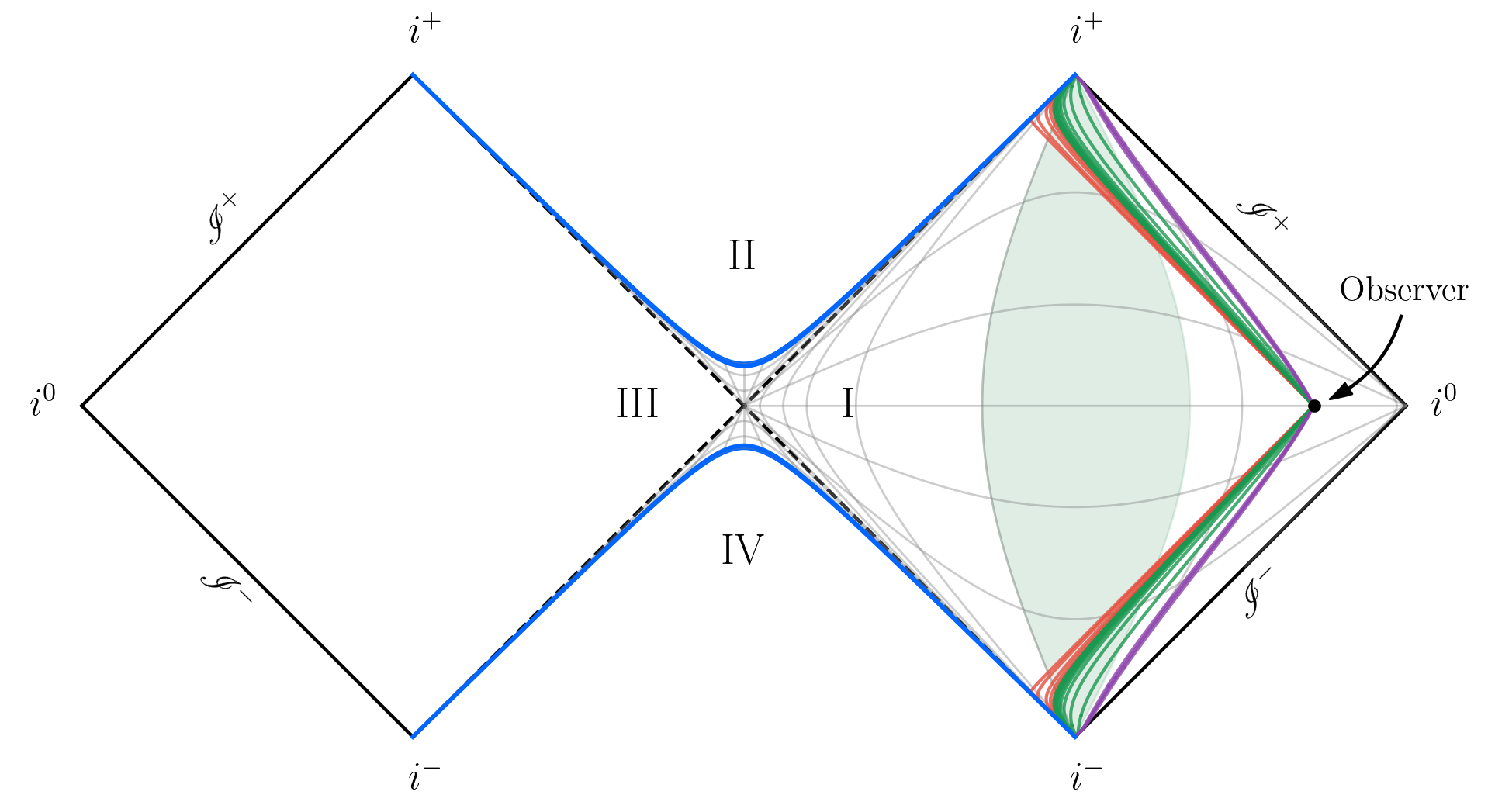}\\[6pt]
\includegraphics[width=0.49\textwidth]{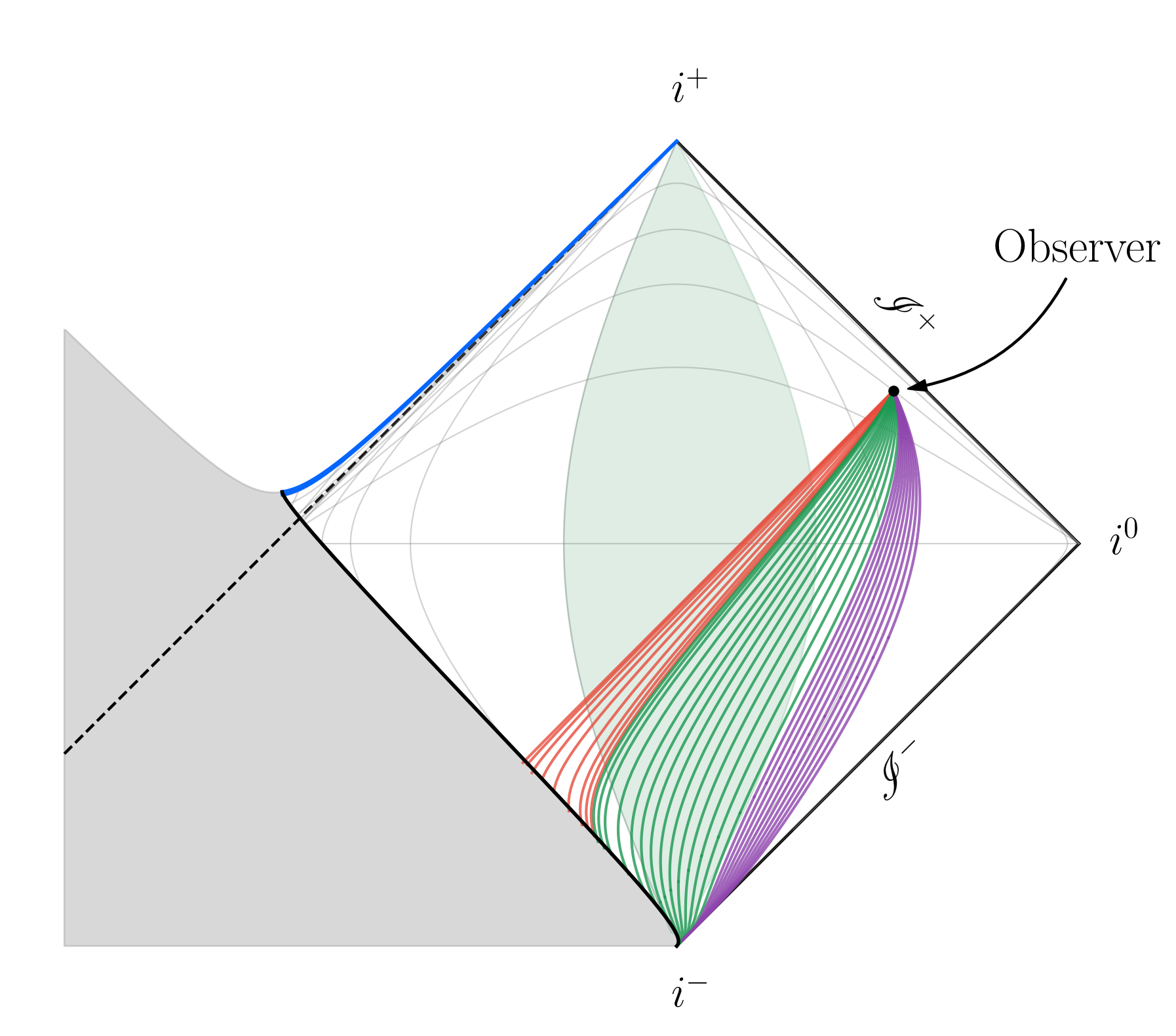}
\caption{Penrose diagrams of the eternal (top) and collapse (bottom) Schwarzschild spacetimes in the rescaled compactification of Eqs.~\eqref{eq:compact_eta}--\eqref{eq:compact_chi} with $S = 10$, with color-coded null geodesics through the location of the observer (black dot). Geodesic colors indicate the three classes discussed in Section~\ref{sec:image}: captured (red), scattered disk-crossing (green), and scattered non-disk-crossing (purple). The green band is the accretion disk, whereas the gray region in the bottom panel is the excised stellar interior, bounded by the surface worldline (bold black curve). The blue lines mark the singularity $r = 0$ and the emitting surface $r = \rmin$, which visually merge at this scale. Dashed lines are the horizons. In the top panel each ray is drawn as a complete geodesic through the observer's location. In the bottom panel only the past-directed portions used by the ray-tracing are shown.}
\label{fig:penrose_eternal}
\end{figure}

The singularity lines, however, do change shape with $S$. From Eqs.~\eqref{eq:schw_from_ks} and~\eqref{eq:compact_inverse}, the condition $r = 0$ is equivalent to $V^2 - U^2 = 1$, or $S^{2}\tan(\eta+\chi)\tan(\eta-\chi) = 1$. Using an identity for the product of tangents, we obtain:
\begin{equation}\label{eq:sing_line}
    \cos 2\eta = \frac{S^{2} - 1}{S^{2} + 1}\,\cos 2\chi .
\end{equation}
At $S = 1$ this yields the straight lines $\eta = \pm\pi/4$ that bound the diamond from above (Region~II) and below (Region~IV). For $S > 1$ the lines sag toward the center of the diamond, as seen in Figure~\ref{fig:penrose_eternal}. Values $0 < S < 1$ would instead magnify the neighborhood of the bifurcation sphere $U = V = 0$, and the singularity lines then bulge outward, approaching null infinity ($\eta = \pi/2 - |\chi|$) as $S \to 0$.

\begin{figure}[htbp]
\centering
\includegraphics[width=\textwidth]{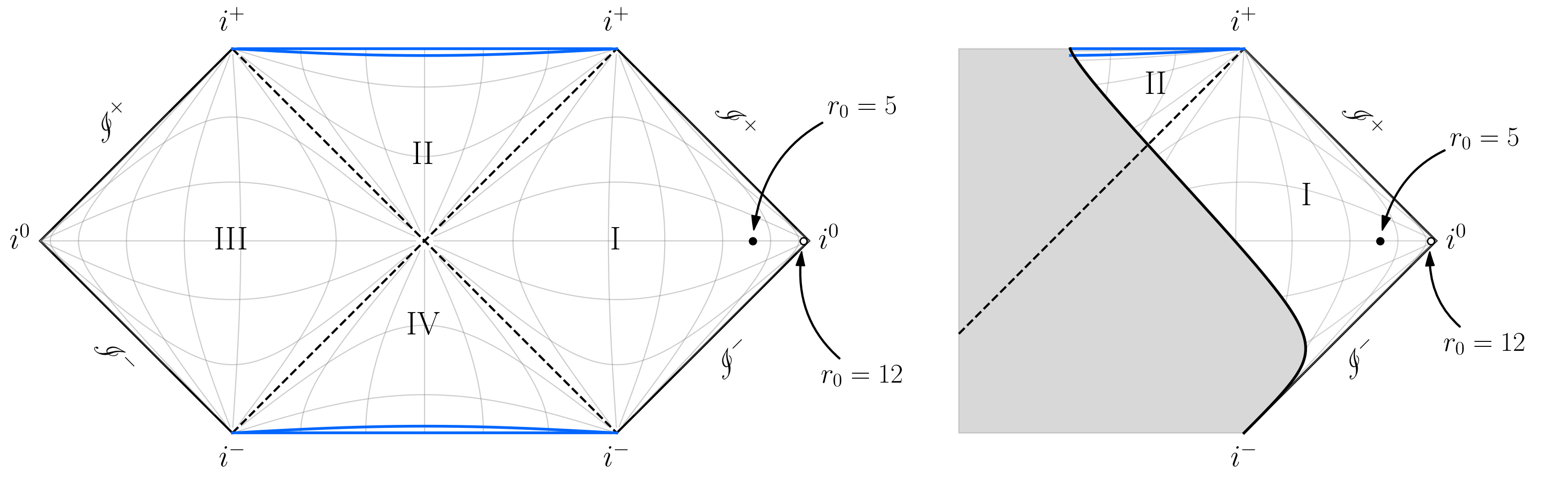}
\caption{Standard Penrose diagrams of the eternal Schwarzschild spacetime (left) and of a spacetime formed by the collapse of a dust ball from rest at $r = 20$ (right), with no accretion disk. Two stationary observers are indicated in each diagram: a close observer at $r_0 = 5$ (filled dot) and a distant observer at $r_0 = 12$ (open dot). The blue lines mark the singularity $r = 0$ and the emitting surface $r = \rmin$. At $S = 1$ (see Eqs.~\eqref{eq:compact_eta}--\eqref{eq:compact_chi}), the distant observer is visually indistinguishable from spatial infinity $i^0$.}
\label{fig:penrose_standard}
\end{figure}

For the astrophysically relevant collapse spacetime, Regions~III and~IV and portions of Regions~I and~II are replaced with the interior of a spherically symmetric collapsing cloud (or star). By Birkhoff's theorem, the remaining portions of Regions~I and~II are identical to the vacuum Schwarzschild case. In practice, we model the collapse by the worldline of the stellar surface and truncate the Schwarzschild geometry at a radial timelike geodesic falling freely from rest at a finite radius. The region swept by the collapsing matter is then excised. Any shading of the excised region in the diagrams below is schematic, as we do not explicitly match the exterior metric to an interior dust solution. Only the surface worldline itself and the segment of the black hole singularity to the future of the surface's endpoint are physical elements of the truncated geometry.

Figure~\ref{fig:penrose_standard} shows both spacetimes side by side in the standard $S = 1$ compactification, with two hypothetical stationary observers marked in each: a close observer at $r_0 = 5$ and a distant observer at $r_0 = 12$. The figure makes the problem apparent: at $S = 1$, even the moderately distant observer at $r_0 = 12$ is visually indistinguishable from spatial infinity $i^0$. The standard compactification is well-adapted to the near-horizon causal structure but heavily compresses the exterior: a stationary observer at Schwarzschild radius $r_0$ sits at Kruskal radius $\sqrt{U^2 + V^2} = \sqrt{r_0/2 - 1}\,e^{r_0/4}$, which is already ${\approx}\,45$ at $r_0 = 12$ and grows exponentially thereafter. After the $\arctan$ of Eqs.~\eqref{eq:compact_eta}--\eqref{eq:compact_chi}, such an observer collapses to within one and a half percent of $\pi/2$, indistinguishable by eye from $i^0$ (Figure~\ref{fig:penrose_standard}).

By contrast, Figure~\ref{fig:penrose_eternal} shows both spacetimes after choosing $S$ well above unity ($S = 10$). This moves the observer visibly inward from null infinity and stretches the exterior region at the expense of the horizon vicinity, without altering any of the canonical boundaries. It now becomes possible to visualize the past-directed light rays that are relevant for ray-tracing the shadow of a black hole. On both diagrams we show representative null geodesics received by an observer at $r_0 = 12$ in the vertical cross-section of the observer's sky (or camera). They span three classes relevant to the ray-tracing. Rays with impact parameter $b$ below the critical value $\bcrit = 3\sqrt{3}$ are captured (red): traced into the past, they cross the past horizon and end on the WH singularity in the eternal spacetime (top panel) or on the collapsing stellar surface in the astrophysical one (bottom panel). Rays with $b > \bcrit$ escape to past null infinity, either crossing the accretion disk on the way (green) or missing it (purple).

Let us now consider the behavior of the two spacetimes with respect to a time translation. If $t_0$ denotes the Schwarzschild time of the observer at $r=r_0$, let us introduce $t' = t - t_0$. Because Schwarzschild time enters Eq.~\eqref{eq:ks_from_schw} only through $\cosh(t/4)$ and $\sinh(t/4)$, the Kruskal--Szekeres coordinates computed from $t'$ are a linear combination of the original ones,
\begin{equation}\label{eq:ks_boost}
    U' = U \cosh(t_0/4) - V \sinh(t_0/4), \qquad
    V' = V \cosh(t_0/4) - U \sinh(t_0/4),
\end{equation}
in both the exterior and the interior charts. This is a Lorentz boost in the $(U, V)$ plane, i.e., a hyperbolic rotation by the angle $t_0/4$~\cite{Landau:1975pou}. The boost leaves $U^2 - V^2$ invariant, and with it every line of constant $r$ (see Eq.~\eqref{eq:schw_from_ks}). In particular, the horizons ($U^2 - V^2 = 0$), the singularities ($U^2 - V^2 = -1$), and null infinity ($U^2 - V^2 \to +\infty$) occupy the same Penrose-diagram loci for any $t_0$. That is, the eternal spacetime is time-translation invariant.

\begin{figure}[htbp]
\centering
\includegraphics[width=\textwidth]{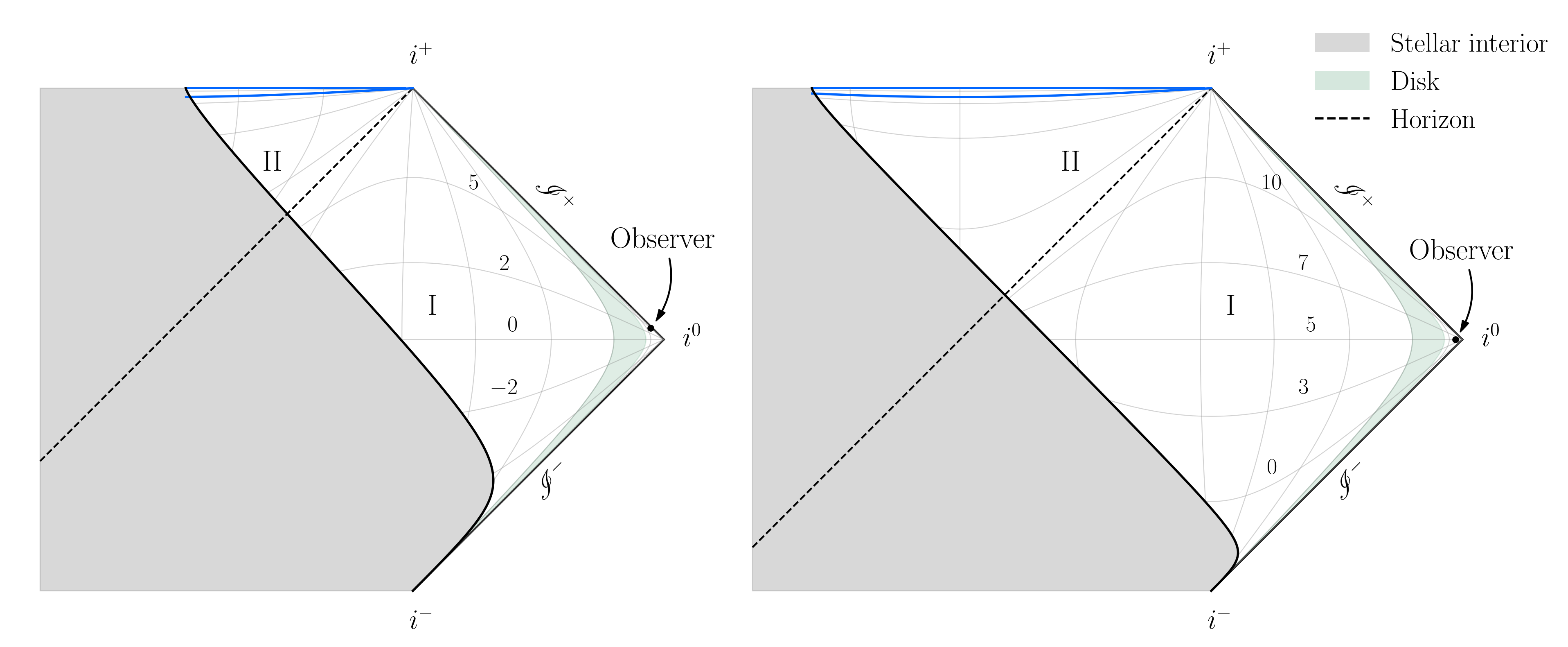}
\caption{Penrose diagrams of the collapse spacetime formed by a dust shell (bold black curve) starting from rest at $r = 20$, both at $S = 1$. Regions~III and~IV are absent, replaced by the stellar interior (gray). \textit{Left:} Standard unshifted compactification. \textit{Right:} Observer-adapted frame (translation $t \to t - t_0$). The collapsing surface is exponentially compressed against the horizon.}
\label{fig:penrose_collapse}
\end{figure}

The collapse spacetime, on the other hand, is not time-translation invariant, as is clear, for example, from the fact that the collapse defines a preferred epoch: the stellar surface starts from rest at a finite radius, which sets an intrinsic timescale for the subsequent collapse\,\footnote{We assume a cloud released from rest at a finite radius. The argument survives in the marginally bound case of release from infinity: the surface sweeping past any fixed radius, such as the observer's, is itself an invariantly distinguished event.}. Figure~\ref{fig:penrose_collapse} shows how the boost of Eq.~\eqref{eq:ks_boost} displaces the worldline of the stellar surface within the diagram while bringing the observer to the center of the diagram. Qualitatively, the timelike worldline of the surface acquires its near-null appearance because the infalling matter accelerates toward the speed of light as $r \to 2$. Therefore, with respect to an observer at late times, the tangent to the worldline approaches a null direction. It is this late portion that is moved into the center of our view by the translation by $t_0$. What was a manifestly timelike curve in the unshifted diagram (cf.\ the bottom panel of Figure~\ref{fig:penrose_eternal}) now hugs the past horizon at close to $45^\circ$, as seen in the right panel of Figure~\ref{fig:penrose_collapse}.

From the practical, ray-tracing standpoint, the two spacetimes are virtually indistinguishable. The progression of Figures~\ref{fig:penrose_standard}--\ref{fig:penrose_collapse} makes it clear that, for an observer far in the future of the collapse, the past geodesic histories in the two geometries coincide up to exponentially small corrections. Rays that escape to $\mathscr{I}^-$ and rays that intersect the equatorial plane before escaping follow identical paths in both. The two spacetimes diverge only in the distant past of the captured rays: in the eternal case these terminate at the WH singularity bounding Region~IV (Figure~\ref{fig:penrose_eternal}, red geodesics in the top panel), while in the collapse case they strike the stellar surface (Figure~\ref{fig:penrose_eternal}, bottom panel). 

According to our assumption that the WH singularity is hidden behind an emitting spacelike surface (Section~\ref{sec:intro}), the rays emerging from under the past horizon of the eternal spacetime produce a glow. The stellar surface emits as well, so the collapse spacetime also produces a glow inside the shadow. This is an image of the surface frozen against the horizon, the so-called ``frozen star''~\cite{zeldovichnovikov1971}. Its optical appearance was analyzed in detail in~\cite{1968ApJ...151..659A,1979ApJ...232..277L} (see~\cite{Yoshino:2019qsh,Ortiz:2015rma,Cao:2025qzy} for modern ray-traced visualizations). This glow, however, is short-lived. It is exponentially redshifted with the observer's time and is in practice visible only to an early observer. The frequency shift and emission location along the captured rays are what distinguish the persistent WH ``bright spot'' from the transient frozen image of the astrophysical stellar surface.

\section{Glow (and shadow) of an eternal black hole}
\label{sec:brightness}

In this section we compute the observer-sky image of the eternal black hole. In Section~\ref{sec:spot} we integrate past-directed null geodesics into the WH interior and obtain the brightness of the glow from the associated frequency shift. We then set up the observer's camera and construct the image of the glow superposed on that of a thin Keplerian accretion disk in Section~\ref{sec:image}. We normalize the observed flux by assuming that both the WH and the disk radiate with unit rest-frame specific intensity. The physically motivated emission models for the two components are deferred to Section~\ref{sec:normalization}.

Let us expand on our assumption (Section~\ref{sec:intro}) that the WH singularity is hidden behind an emitting spacelike surface. In Markov's terms, this surface bounds the region where the density of matter would reach its limiting value. Inside the event horizon of a Schwarzschild black hole the radial coordinate $r$ is timelike. Accordingly, the geometry described by Eq.~\eqref{eq:schw_metric} is not a static gravitational field but an anisotropic cosmology, or a T-region~\cite{2001GReGr..33.2259N}, with hypersurfaces of constant $r$ playing the role of successive moments in time. Near the singularity the metric approaches a Kasner-type regime (see, for example,~\cite{Landau:1975pou,Zeldovich:1983cr}). The WH Region~IV of Figure~\ref{fig:penrose_eternal}, in particular, emerges from a past spacelike singularity at $r = 0$ and expands toward the past horizon at $r = 2$, with its spatial sections decreasing in curvature as $r$ grows --- much like the universe does after the Big Bang. By analogy with cosmological recombination~\cite{Zeldovich:1983cr}, we assume that the matter streaming out of the singularity is opaque at high curvature and becomes transparent only at some ``surface of last scattering'' $r = \rmin$. In the same cosmological spirit, we also assume a source (emitter) that is \textit{comoving} at $r=\rmin$.

We leave $\rmin$ as a free parameter throughout this section (see Section~\ref{sec:normalization} for a scenario with Planck-scale decoupling). Note that the nature of the matter and possible quantum effects that might lead to its production are beyond the scope of this paper.

\subsection{White hole bright spot}
\label{sec:spot}

Before we can construct the observer-sky image of the WH spot, we need to integrate null geodesics past-directed from the observer through the past horizon and into Region~IV, identifying those that terminate on the surface of last scattering $r = \rmin$. The Kruskal--Szekeres coordinates of Section~\ref{sec:spacetime} cover all four regions at once, but they are poorly suited to numerical integration. To begin with, $r$ is defined only implicitly, requiring an evaluation of the Lambert $W$ function at every step. Another issue is the exponential growth of $U$ and $V$, which leads to floating-point overflow once $r \pm t$ exceeds a few thousand, while near the horizons $U^{2} - V^{2}$ degrades into a difference of huge, nearly equal numbers.

We therefore trade $U$ and $V$ for the outgoing and ingoing Eddington--Finkelstein times, defined in terms of the tortoise coordinate $r_\star$,
\begin{equation}\label{eq:ef_def}
    \bar{\eta} = t - r_\star, \qquad \bar{\chi} = t + r_\star, \qquad r_\star(r) = r + 2\,\ln\!\left|\frac{r}{2} - 1\right|,
\end{equation}
which are regular across the past and future horizons, respectively: a single Eddington--Finkelstein chart covers two regions of the Penrose diagram at once\,\footnote{In terms of the Kruskal--Szekeres coordinates of Eq.~\eqref{eq:ks_from_schw}, $U - V = e^{-\bar{\eta}/4}$ and $U + V = e^{\bar{\chi}/4}$. The outgoing chart is therefore the wedge $U - V > 0$: Regions~I and IV are connected along the past horizon, with the future horizon pushed to $\bar{\eta} \to +\infty$. Likewise, the ingoing chart is the wedge $U + V > 0$, covering Regions~I and II.}. Our rays cross the past horizon, so we work in the outgoing chart $(\bar{\eta}, r, \theta, \varphi)$, in which the metric reads
\begin{equation}\label{eq:ef_metric}
    ds^{2} = \left(1 - \frac{2}{r}\right) \dd \bar{\eta}^{2} + 2\, \dd \bar{\eta}\, \dd r - r^{2}\, \dd\Omega^{2}\,.
\end{equation}
A photon carries the 4-momentum $p^{\mu} \equiv \dot{x}^{\mu}$, where the dot stands for a derivative with respect to the affine parameter $\lambda$ along the ray. The metric~\eqref{eq:ef_metric} does not depend on $\bar{\eta}$, so $\xi \equiv \partial_{\bar{\eta}}$ is a Killing vector: it coincides with $\partial_{t}$ where both charts apply and generates the time translation of Eq.~\eqref{eq:ks_boost}. It is timelike in Region~I (the exterior is static) and spacelike in Region~IV, where it generates spatial translations along the cylindrical sections $\mathbb{R} \times S^{2}$ of the interior cosmology. Either way, $E \equiv p_{\mu}\xi^{\mu} = p_{\bar{\eta}}$ is conserved along geodesics in both regions and smooth across the past horizon. Spherical symmetry confines the ray to a plane and conserves $L \equiv -p_{\Phi}$, with $\Phi$ the azimuthal angle in that plane. To within a rescaling of the affine parameter, a null geodesic depends on the two constants only through the impact parameter $b \equiv L/E$.

With $\lambda$ rescaled so that $E = 1$, the conservation laws and the null condition $p_{\mu}p^{\mu} = 0$ yield the equations of motion:
\begin{align}
    \dot{r} &= w\,, \label{eq:eom_r}\\
    \dot{w} &= \frac{b^{2}}{r^{3}}\!\left(1 - \frac{3}{r}\right), \label{eq:eom_w}\\
    \dot{\Phi} &= \frac{b}{r^{2}}\,, \label{eq:eom_Phi}\\
    \dot{\bar{\eta}} &= \frac{b^{2}}{r^{2}\,(1 + w)} \qquad (\bar{\eta} \to \bar{\chi},\ w \to -w\ \text{in the ingoing chart})\,, \label{eq:eom_eta}
\end{align}
where
\begin{equation}
    w \equiv \pm\sqrt{R}\,, \qquad
    R(r, b) \equiv 1 - \frac{b^{2}}{r^{2}}\!\left(1 - \frac{2}{r}\right)\,,
\end{equation}
is an auxiliary function that handles radial turning points naturally: there $w$ vanishes while $\dot{w}$ stays finite, so the sign flips smoothly and no branch switching of the square root is needed. Eqs.~\eqref{eq:eom_r}--\eqref{eq:eom_eta} are used to obtain the full geodesics depicted in Figure~\ref{fig:penrose_eternal}. The ratio of Eqs.~\eqref{eq:eom_r} and~\eqref{eq:eom_Phi} yields the usual equation for the photon orbits:
\begin{equation}\label{eq:geodesic_orbit}
    \frac{\dd r}{\dd\Phi} = \pm\, \frac{r^{2}}{b}\,\sqrt{R(r, b)}\,,
\end{equation}
which leads to the textbook classification of null rays. For captured rays, $b < \bcrit \equiv 3\sqrt{3}$, the radial potential $R(r,b)$ stays positive along the entire trajectory (its minimum, reached at the photon sphere $r = 3$, is $1 - b^{2}/27$). In particular, a captured ray extended into the past falls monotonically from the observer radius $r_0$, crosses the past horizon into Region~IV, and terminates on the surface of last scattering for any $\rmin < 2$. Rays with $b > \bcrit$ instead reach a radial turning point, $w = R = 0$ at some $r > 3$, and swing back to infinity.

To calculate the brightness of the spot, we in fact need only Eq.~\eqref{eq:geodesic_orbit}. Indeed, the shift in frequency between a source and an observer is determined from the factor
\begin{equation}\label{eq:g_def}
    g \equiv \frac{\nu_{\mathrm{obs}}}{\nu_{\mathrm{emit}}} =  \frac{(p_\mu u^\mu)_{\mathrm{obs}}}{(p_\mu u^\mu)_{\mathrm{emit}}}\,,
\end{equation}
where $u^\mu$ is the 4-velocity of the source or observer. By the Lorentz invariance of $I_\nu/\nu^{3}$~\cite{Misner:1973prb}, the observed specific intensity is
\begin{equation}\label{eq:I_obs_unit}
    I_\nu^{\mathrm{obs}} = g^{3}\, I_\nu^{\mathrm{emit}} = g^{3}\,,
\end{equation}
where we normalized to unit intensity at emission.

By our assumption, the matter covering the WH singularity is comoving with the interior cosmology. Its worldlines are the radial timelike geodesics of constant $(t, \theta, \varphi)$, and the surface of last scattering $r = \rmin$ is a hypersurface of constant cosmological time. In the local Schwarzschild coordinates $(t, r, \theta, \Phi)$ of Region~IV, the comoving 4-velocity has a nonvanishing component only along the timelike coordinate $r$. The normalization $u_{\mu}u^{\mu} = 1$ then relates $r$ to the proper time, $\dd\tau = \dd r/\sqrt{2/r - 1}$, such that $u^{r} = \sqrt{2/r - 1}$. Since $t$ is constant along the worldline, Eq.~\eqref{eq:ef_def} gives $\dd\bar{\eta} = -\dd r_\star = -\dd r\,(1 - 2/r)^{-1}$, and hence
\begin{equation}\label{eq:u_comoving}
\begin{aligned}
    u^{\bar{\eta}} &= -\frac{u^{r}}{1 - 2/r} = \frac{1}{\sqrt{2/r - 1}}\,, \\
    u_{\bar{\eta}} &= \left(1 - 2/r\right) u^{\bar{\eta}} + u^{r} = 0\,, \qquad u_{r} = u^{\bar{\eta}}\,.
\end{aligned}
\end{equation}
Using standard expressions for $\left.u^{\mu}\right|_{\mathrm{obs}}$ and reading the photon 4-momentum off Eqs.~\eqref{eq:eom_r}--\eqref{eq:eom_eta}, we obtain the contractions
\begin{equation}\label{eq:contractions}
\begin{aligned}
    (p_{\mu} u^{\mu})_{\mathrm{emit}} &= p^{r} u_{r} = \frac{\sqrt{R(\rmin, b)}}{\sqrt{2/\rmin - 1}}\,, \\
    (p_{\mu} u^{\mu})_{\mathrm{obs}} &= p_{\bar{\eta}}\, u^{\bar{\eta}} = \frac{1}{\sqrt{1 - 2/r_{0}}}\,,
\end{aligned}
\end{equation}
which result in
\begin{equation}\label{eq:g_spot_body}
    g_{\mathrm{spot}}(b) = \frac{1}{\sqrt{1 - 2/r_0}}\,\left[\frac{1}{2/\rmin - 1} + \frac{b^{2}}{\rmin^{2}}\right]^{-1/2}\,.
\end{equation}

Note that, although the spacetime in Region~IV is anisotropic, the frequency-shift factor depends on the impact parameter alone. This makes the WH spot appear azimuthally symmetric on the observer's sky (cf.~\cite{1968ApJ...151..659A}). In the limit $\rmin \to 2$, $g_{\mathrm{spot}} \sim \sqrt{2/\rmin - 1} \to 0$, i.e., the emission is infinitely redshifted regardless of the impact parameter. One interpretation is that the expansion of the interior halts at the horizon ($u^{r} \to 0$). And as $\rmin \to 0$, $g_{\mathrm{spot}} \sim \rmin/b$ for a distant observer: a significant Doppler boost for $b \ll \rmin$ turns into a redshift for $b \gtrsim \rmin$, down to $g_{\mathrm{spot}} \sim \rmin/\bcrit \ll 1$ at the edge of the shadow. More precisely, at the center of the shadow ($b = 0$),
\begin{equation}\label{eq:g_spot_central}
    g_{\mathrm{spot}}(0) = \sqrt{\frac{2/\rmin - 1}{1 - 2/r_0}}\,,
\end{equation}
which exceeds unity for $\rmin \lesssim 1$, so the central core is blueshifted. Combining Eq.~\eqref{eq:g_spot_body} with Eq.~\eqref{eq:I_obs_unit} gives the observer-sky intensity of the spot,
\begin{equation}\label{eq:I_spot_body}
    I_\nu^{\mathrm{obs,\,spot}}(b) = g_{\mathrm{spot}}^{3}(b)\;\Theta(\bcrit - b),
\end{equation}
where the Heaviside step function $\Theta$ restricts the emission to captured rays. The spot therefore fills the entire shadow (the $b < \bcrit$ region that in the astrophysical collapse geometry would be perfectly dark) with a radially symmetric bright pattern, peaked at the center and falling off as $(\rmin/b)^{3}$ as $b \to \bcrit$.

\begin{figure}[t]
\centering
\includegraphics[width=0.7\textwidth]{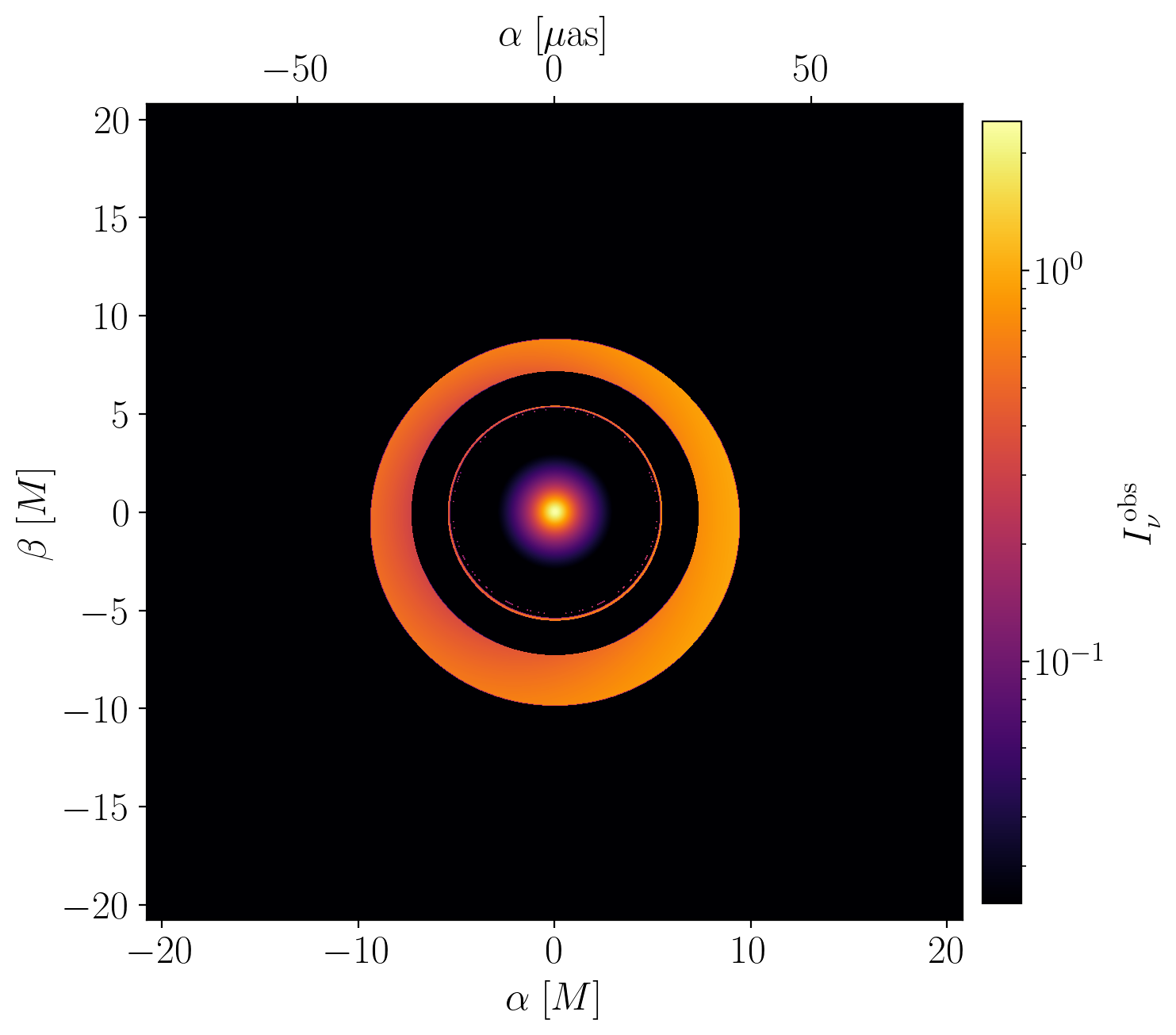}
\caption{Observer-sky image of the eternal Schwarzschild black/white hole with a thin Keplerian accretion disk, at inclination $\iota = 17^{\circ}$. The upper horizontal axis shows the angular scale for M87$^\star$ parameters (Eq.~\ref{eq:uas_per_pixel}).}
\label{fig:observer_image}
\end{figure}

\subsection{Observer-sky image}
\label{sec:image}

To embed the spot in an astrophysical setting we superpose a geometrically thin, optically thick Keplerian accretion disk in the equatorial plane~\cite{Luminet:1979nyg}. We then launch null rays from the observer and integrate them into the past, so that each pixel on the observer's sky is traced back either to a matter source (accretion disk, stellar surface, or WH interior) or to null infinity. The observer is static at radius $r_0$ and inclination $\iota$ (measured from the disk normal). Following the camera setup of~\cite{James:2015yla}, we introduce the spherical polar coordinates $(\delta,\psi)$, where $\delta$ is the angular distance from the optical axis (pointed toward the black hole), and $\psi$ is the position angle around it, measured from the sky projection of the disk normal. A past-directed ray launched at angle $\delta$ from the axis has the impact parameter~\cite{Synge:1966okc,chandrasekhar1983}
\begin{equation}\label{eq:camera}
    b = \frac{r_0}{\sqrt{1 - 2/r_0}}\;\sin\delta\,.
\end{equation}
For an astronomically distant observer, $r_{0}$ is simply the distance to the source, and Eq.~\eqref{eq:camera} reduces to the small-angle form $b \approx r_{0}\,\delta$. The image is rendered in the scaled image-plane coordinates
\begin{equation}\label{eq:ab}
\begin{aligned}
    \alpha &=  b\,\sin\psi\,, & \qquad \beta &=  b\,\cos\psi\,, \\
    b &=  \sqrt{\alpha^{2} + \beta^{2}}\,, & \qquad \psi &=  \arctan_{2}(\alpha, \beta)\,,
\end{aligned}
\end{equation}
with the $\beta$ axis along the sky projection of the disk normal and the $\alpha$ axis along the local direction of increasing azimuth $\varphi$. Here $\arctan_{2}(y, x)$ denotes the two-argument arctangent, which returns the angle in $(-\pi, \pi]$ with the quadrant fixed by the signs of both arguments. Unlike the principal-branch $\arctan(y/x)$, it is single-valued everywhere, including $x = 0$.

We use the standard circular Keplerian disk with 4-velocity~\cite{novikov1973}
\begin{equation}\label{eq:u_disk}
    u_{\mathrm{disk}}^{\mu} = \left(1 - \frac{3}{r}\right)^{\!-1/2}\,\bigl(1,\; 0,\; 0,\; \Omega(r)\bigr), \qquad \Omega(r) = r^{-3/2}\,,
\end{equation}
and the frequency shift of Eq.~\eqref{eq:g_def} for a ray that hits the disk at radius $r$,
\begin{equation}\label{eq:g_disk_body}
    g_{\mathrm{disk}}(r, b_\varphi) = \frac{(p_\mu u^\mu)_{\mathrm{obs}}}{p_\mu u_{\mathrm{disk}}^{\mu}} = \frac{1}{\sqrt{1 - 2/r_0}}\;\frac{\sqrt{1 - 3/r}}{1 - b_\varphi\,\Omega(r)}\,, \qquad b_\varphi = \alpha\sin\iota\,.
\end{equation}

Due to the spherical symmetry, a trajectory depends on the position on the observer's sky only through the impact parameter $b$. We therefore integrate the orbit equation~\eqref{eq:geodesic_orbit} only once, for a one-dimensional grid of impact parameters refined near $\bcrit$, and store the resulting library of trajectories $r(\Phi; b)$. The position angle $\psi$ enters only through the azimuths at which a ray pierces the equatorial plane, $\Phi^{(n)} = \Phi_{0} + n\pi$ ($n = 0, 1, 2, \ldots$), where
\begin{equation}\label{eq:Phi0_body}
    \Phi_{0} = \arctan_{2}(\cot\iota,\; -\cos\psi) \in (0, \pi)
\end{equation}
is the azimuth of the first crossing and follows from elementary spherical trigonometry on the celestial sphere. Each crossing that falls inside the disk annulus, $\rin \le r_n \le \rout$ with $r_n \equiv r(\Phi^{(n)}; b)$, contributes to the observed intensity. Combining Eq.~\eqref{eq:g_disk_body} with Eq.~\eqref{eq:I_obs_unit},
\begin{equation}\label{eq:I_disk_body}
\begin{aligned}
    I_\nu^{\mathrm{obs,\,disk}}(\alpha, \beta) &=  \sum_{n\,\ge\,0} g_{\mathrm{disk}}^{3}\!\bigl(r_n,\, b_\varphi\bigr)\,, \\
    I_\nu^{\mathrm{obs}} &=  I_\nu^{\mathrm{obs,\,spot}} + I_\nu^{\mathrm{obs,\,disk}}\,.
\end{aligned}
\end{equation}
Because the disk is opaque, higher-order crossings at pixels already covered by the direct image do not contribute. Successive $n$ produce the direct image ($n = 0$), the primary photon ring ($n = 1$), and exponentially thinner higher-order rings that cluster toward the critical curve $b = \bcrit$.

Figure~\ref{fig:observer_image} shows the resulting image for the parameters $r_{0} = 12$, $\rmin = 0.8$, $\rin = 6$ (the innermost stable circular orbit), $\rout = 9$, and $\iota = 17^{\circ}$. The inclination value is motivated by the value adopted for M87$^\star$ by the EHT collaboration~\cite{CraigWalker:2018vam,EventHorizonTelescope:2019pcy}. Recall that the geometry throughout this paper is strictly Schwarzschild, and we use this specific inclination for illustration only. For the same reason, the upper horizontal axis of Figure~\ref{fig:observer_image} is calibrated in microarcseconds assuming M87$^\star$ parameters $M = 6.5 \times 10^{9}\,M_{\odot}$ and $r_{0} = 16.8\,\mathrm{Mpc}$~\cite{EventHorizonTelescope:2019dse}. The image is computed at a resolution of $N = 1024$ pixels per side, and the per-pixel angular scale for a side length $L$ in units of $M$ is $\Delta\alpha = (L/N)\,GM/(c^{2}r_{0})$. Evaluated at the full width $L = 8\,\bcrit \approx 41.57$ of Figure~\ref{fig:observer_image}, this reads (see also Eq.~4 in~\cite{Popov:2021aru}):
\begin{equation}\label{eq:uas_per_pixel}
    \Delta\alpha \approx 0.155\;\frac{\mu\mathrm{as}}{\mathrm{pixel}}\;\left(\frac{1024}{N}\right)\!\left(\frac{M}{6.5\times 10^{9}\,M_{\odot}}\right)\!\left(\frac{r_{0}}{16.8\,\mathrm{Mpc}}\right)^{\!-1}.
\end{equation}

Inside the critical curve $b = \bcrit$, the WH spot fills the shadow with the radially symmetric pattern that is bright in the center and becomes fainter toward the edge (see Eq.~\eqref{eq:g_spot_body}). Outside the critical curve, the direct disk image ($n = 0$) exhibits a faint Doppler asymmetry barely visible at this near-face-on inclination, and the primary photon ring ($n = 1$) appears as a thinner ring just outside $\bcrit$\,\footnote{At our working resolution of $1024 \times 1024$ pixels we also trace 74 pixels of the secondary ring ($n = 2$), visible in Figure~\ref{fig:observer_image} with sufficient zoom. We make no effort to bring them out, as the higher-order rings play no role in the analysis that follows.}. Both sources are plotted on a common logarithmic color scale and assume unit rest-frame emissivity. The normalized interferometric visibilities are analyzed in Section~\ref{sec:visibility}, including examples of the disk/spot flux ratio in Section~\ref{sec:normalization}.

\section{Interferometric signatures}
\label{sec:visibility}

The basic observable of an interferometer is the complex visibility of the source. In the idealized setting adopted here (no instrumental noise, calibration artifacts, etc.), the visibility is given by the Fourier transform that follows from the van Cittert--Zernike theorem~\cite{thompson2017},
\begin{equation}\label{eq:vis}
    V(\mathbf{q}) = \int I(\mathbf{x})\, e^{-2\pi i\, \mathbf{q}\cdot\mathbf{x}}\, \dd^{2}\mathbf{x},
\end{equation}
where $\mathbf{x} = (\alpha, \beta)$ are the image-plane coordinates of Section~\ref{sec:image} (in units of $M$) and $\mathbf{q} = (u, v)$ is the conjugate baseline vector (in units of $1/M$). The observational convention is instead to measure baseline lengths in units of the observing wavelength $\lambda$~\cite{thompson2017}. Indeed, observations deal with dimensionless angles on the sky, $\approx \mathbf{x}/r_{0}$ (Eq.~\eqref{eq:camera}), which are resolved down to the smallest angle $\sim \lambda/D$, where $D$ is the projected baseline length. The conjugate baseline vector is then measured in units of $D/\lambda$. With the M87$^\star$ fiducials of Section~\ref{sec:image}, the Fourier-plane pixel (assuming no zero-padding) is
\begin{equation}\label{eq:baseline_Glambda}
    \Delta u = \frac{1}{N\,\Delta\alpha} \approx 1.30\;\frac{\mathrm{G}\lambda}{\mathrm{pixel}}\;\left(\frac{M}{6.5\times 10^{9}\,M_{\odot}}\right)^{\!-1}\!\left(\frac{r_{0}}{16.8\,\mathrm{Mpc}}\right)\,.
\end{equation}
Note that $\Delta u$ is independent of the resolution $N$. At a constant field of view $L$, lower resolution results in a narrower range of baselines as expected. For comparison, the longest EHT baselines, $\approx 8\,\mathrm{G}\lambda$~\cite{EventHorizonTelescope:2019dse}, fall at $q \approx 0.15\,M^{-1}$ at the M87$^\star$ mass and distance.

The EHT presented its evidence for the M87$^\star$ and Sgr~A$^\star$ shadows as raw visibility amplitudes plotted against projected baseline length~\cite{EventHorizonTelescope:2019dse,EventHorizonTelescope:2022wkp}. As a proxy for those plots, here we use the azimuthally averaged squared visibility
\begin{equation}
\langle|V|^{2}\rangle(q) \equiv (2\pi)^{-1}\int_{0}^{2\pi}|V(q,\phi)|^{2}\,\dd\phi\,,
\end{equation}
where $(q, \phi)$ are polar coordinates in the Fourier plane following the convention of~\cite{thompson2017}, with $\phi$ measured from the same direction as the image position angle $\psi$. As we will see below, this averaged form allows for a closed-form, analytical expression for the WH spot contribution. More generally, an arbitrary intensity distribution in the image plane can be expanded in the Fourier series $I(b, \psi) = \sum_{m} I_{m}(b)\,e^{im\psi}$. Using the Jacobi--Anger identity, this yields
\begin{equation}\label{eq:vis_modes}
    V(q, \phi) = 2\pi \sum_{m} (-i)^{m}\, \tilde{I}_{m}(q)\, e^{im\phi}, \qquad
    \tilde{I}_{m}(q) = \int_{0}^{\infty} I_{m}(b)\, J_{m}(2\pi qb)\, b\, \dd b\,,
\end{equation}
where $\tilde{I}_{m}$ is the $m$-th order Hankel transform of the $m$-th azimuthal mode. Squaring and averaging over $\phi$ makes the cross terms vanish by orthogonality, giving the Parseval-type identity
\begin{equation}\label{eq:vis2_avg}
    \langle |V|^{2} \rangle(q) = (2\pi)^{2} \sum_{m=-\infty}^{\infty} |\tilde{I}_{m}(q)|^{2}\,.
\end{equation}

In Figure~\ref{fig:vis_per_feature} we plot the root-mean-square $\sqrt{\langle|V|^{2}\rangle}$, which gives an upper bound on the azimuthally averaged visibility magnitude, $\langle|V|\rangle \le \sqrt{\langle|V|^{2}\rangle}$ (Cauchy--Schwarz). For an axisymmetric image only $m = 0$ survives and $\sqrt{\langle|V|^{2}\rangle} = 2\pi|\tilde{I}_{0}(q)|$ exactly. In particular, this applies to the WH spot and, to a lesser degree, to the disk image and the photon ring. The sources of non-axisymmetric power are the distortion due to projection on the image plane and the Doppler asymmetry of the inclined disk, both suppressed at the M87-motivated inclination $\iota = 17^{\circ}$.

In what follows we evaluate the visibility of each of the three features in closed form with normalized emissivity and show their characteristic baseline-length scalings (Section~\ref{sec:signatures}). We then restore physical normalization by computing the WH spot flux in the EHT band and show constraints on it as a function of~$\rmin$ and emission temperature~$T$ (Section~\ref{sec:normalization}).

\subsection{Three signatures with normalized brightness}
\label{sec:signatures}

Here we compute the factor $g^3$ for the three features: WH spot, disk, and the photon ring. This is equivalent to the assumption of unit specific intensity at emission. At the near-face-on inclination, each feature is approximately axisymmetric, so we keep only the $m = 0$ mode of Eq.~\eqref{eq:vis_modes}. We begin with the WH spot, which admits an exact closed form, and then briefly treat the disk and the photon ring, whose scalings with~$q$ serve as consistency checks.

Let us recast the bright spot profile of Eqs.~\eqref{eq:I_spot_body} and~\eqref{eq:g_spot_body} as follows:
\begin{equation}\label{eq:spot_rational}
    I_{\mathrm{spot}}(b) = \frac{C^{3}\,\rmin^{3}}{(b^{2} + z^{2})^{3/2}}\,\Theta(\bcrit - b),
\end{equation}
with $A = (2/\rmin - 1)^{-1}$, $C = (1 - 2/r_{0})^{-1/2}$, and $z = \rmin\sqrt{A} = \rmin/\sqrt{2/\rmin - 1}$. If we extend this profile beyond $\bcrit$ to infinity, using the Hankel pair~\cite{piessens2000}
\begin{equation}\label{eq:hankel_exp}
    \int_{0}^{\infty} \frac{J_{0}(sb)\,b\,\dd b}{(b^{2} + z^{2})^{3/2}} = \frac{e^{-sz}}{z}\,,
\end{equation}
we obtain a simple exponential in the baseline length:
\begin{equation}\label{eq:V_spot_analytic}
    V_{\infty}(q) = \frac{2\pi C^{3}\,\rmin^{3}}{z}\; e^{-2\pi q z}.
\end{equation}
The decay rate $2\pi z = 2\pi\rmin / \sqrt{2/\rmin - 1}$ is set entirely by the emission radius $\rmin$ inside the WH. The zero-baseline amplitude is $V_{\infty}(0) = 2\pi C^{3}\rmin^{2}\sqrt{2/\rmin - 1} = 2\pi z^{2} g_{\mathrm{spot}}^{3}(0)$, the central intensity times the effective area of the core of radius $z$. For $\rmin \ll 2$ it scales as $\rmin^{3/2}$, so a smaller last-scattering radius makes the spot both fainter and flatter in the $uv$-plane. Interestingly, the visibility displays no oscillatory behavior in the $uv$-plane but rather decays monotonically. This reflects a general property of the Fourier transform: the smoother the profile, the faster and smoother the decay of its transform~\cite{Iosevich:2014}. Sharp edges, by contrast, produce slowly decaying oscillatory transforms, as the disk and the photon ring illustrate below.

\begin{figure}[!htbp]
\centering
\includegraphics[width=0.8\textwidth]{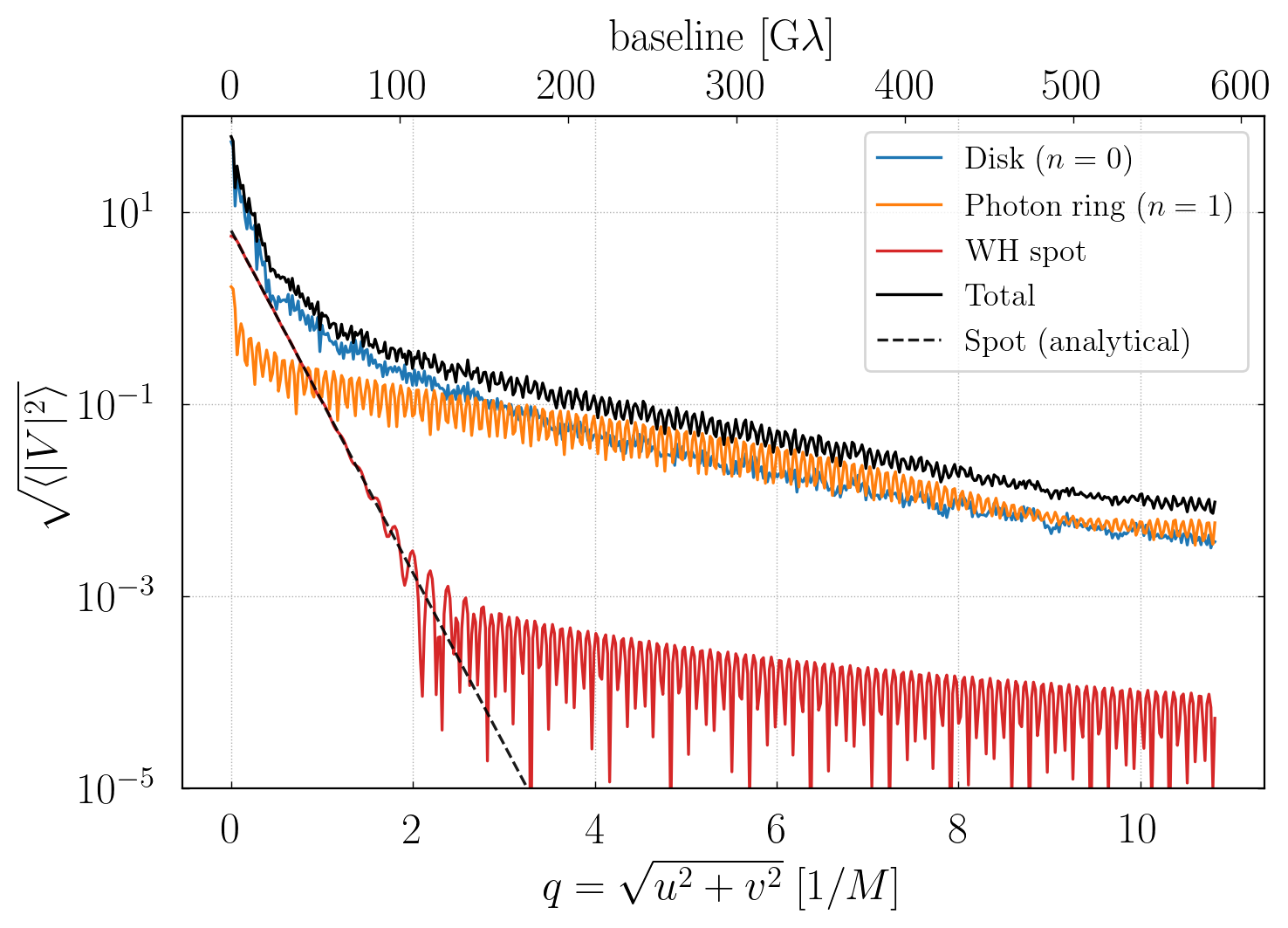}
\caption{RMS visibility amplitude for three features: disk (blue), photon ring (orange), WH spot (red), and total (black), for the value $\rmin = 0.8$ (same as in the previous sections). The dashed line is the analytical expression~\eqref{eq:V_spot_analytic}, with quasi-periodic truncation corrections, Eq.~\eqref{eq:V_tail_asym}, appearing beyond $q \approx 1.5$. The upper horizontal axis converts $q$ to the physical baseline length for the M87$^\star$ parameters (cf.\ Eq.~\eqref{eq:baseline_Glambda}).}
\label{fig:vis_per_feature}
\end{figure}

The truncation at $b = \bcrit$ replaces $V_{\infty}$ with $V_{\infty} - V_{\mathrm{tail}}$, where $V_{\mathrm{tail}}$ is the contribution of the profile from $b > \bcrit$. Integrating by parts twice, we obtain the asymptotic expansion:
\begin{equation}\label{eq:V_tail_asym}
    V_{\mathrm{tail}}(q) = -\frac{\bcrit\, g_{\mathrm{spot}}^{3}(\bcrit)}{q}\, J_{1}(2\pi q \bcrit) - \frac{\bcrit\, \bigl(g_{\mathrm{spot}}^{3}\bigr)'(\bcrit)}{2\pi q^{2}}\, J_{0}(2\pi q \bcrit) + \mathcal{O}(q^{-7/2})\,.
\end{equation}
The corrections are quasi-periodic with period $\Delta q \sim 1/\bcrit$, set by the sharp photon-sphere cutoff, and the first-order term decays only as $q^{-3/2}$. Their amplitude, however, is small. It is proportional to $g_{\mathrm{spot}}^{3}(\bcrit) = g_{\mathrm{spot}}^{3}(0)\,[z^{2}/(z^{2} + \bcrit^{2})]^{3/2} \approx g_{\mathrm{spot}}^{3}(0)\,(z/\bcrit)^{3}$, which is $\approx 2\times10^{-3}\,g_{\mathrm{spot}}^{3}(0)$ at the $\rmin = 0.8$ of the figures. The tail therefore takes over where the exponential~\eqref{eq:V_spot_analytic} falls below this level, at $q \sim 3\ln(\bcrit/z)/(2\pi z) \approx 1.5$. Though small, they indeed become noticeable in Figure~\ref{fig:vis_per_feature} close to this scale. Below that point the main, analytical term~\eqref{eq:V_spot_analytic} closely tracks the numerical Hankel transform.

As a consistency check, we also reproduce the interferometric signature of the disk and that of the photon ring (Schwarzschild limit) extensively studied in the literature~\cite{Johnson:2019ljv,Gralla:2020yvo,Cardenas-Avendano:2023dzo,Lupsasca:2024xhq}. The disk image is well-modeled by a uniform annulus with inner radius $b_{\mathrm{in}}$, outer radius $b_{\mathrm{out}}$, and total flux $I_{\mathrm{disk,tot}}$: $I_{\mathrm{disk}}(b) = I_{\mathrm{disk,tot}}\,[\pi(b_{\mathrm{out}}^{2} - b_{\mathrm{in}}^{2})]^{-1}$ for $b_{\mathrm{in}} \le b \le b_{\mathrm{out}}$. Its zeroth-order Hankel transform reads
\begin{equation}\label{eq:V_disk}
    V_{\mathrm{disk}}(q) = I_{\mathrm{disk,tot}}\,\frac{b_{\mathrm{out}}\, J_{1}(2\pi qb_{\mathrm{out}}) - b_{\mathrm{in}}\, J_{1}(2\pi qb_{\mathrm{in}})}{\pi\, q\,(b_{\mathrm{out}}^{2} - b_{\mathrm{in}}^{2})}\sim \frac{1}{q^{3/2}} \quad \text{as} \quad q\gg\frac{1}{b_{\mathrm{in}}}\,.
\end{equation}

The photon ring at $n=1$ and higher orders is exponentially thinner than the disk and accumulates toward the critical curve $b = \bcrit$. It is thus well-modeled by a delta function at the critical curve with total flux $I_{\mathrm{ring,tot}}$, $I_{\mathrm{ring}}(b) = I_{\mathrm{ring,tot}}\,\delta(b - \bcrit)/(2\pi\bcrit)$, whose zeroth-order Hankel transform reads
\begin{equation}\label{eq:V_ring}
    V_{\mathrm{ring}}(q) = I_{\mathrm{ring,tot}}\, J_{0}(2\pi q\bcrit)\sim\frac{1}{q^{1/2}} \quad \text{as} \quad q\gg \frac{1}{\bcrit}\sim \frac{1}{b_{\mathrm{in}}}\,. 
\end{equation}
Owing to the slower decay, the ring overpowers the disk at sufficiently long baselines even for comparable total fluxes, as first pointed out in~\cite{Johnson:2019ljv}.

Figure~\ref{fig:vis_per_feature} shows the resulting visibility amplitudes for each feature. The total (black) is dominated by the disk (blue) at relatively short baselines $q$ and overtaken by the photon ring at longer baselines. This is consistent with the slower decay of the ring's visibility ($\sim q^{-1/2}$) as compared to the disk ($\sim q^{-3/2}$). For the fiducial value $\rmin = 0.8$, the WH spot contribution (red) stays below both the disk and ring envelopes across the shown baselines. Note, however, the trade-off between the zero-baseline amplitude and the decay rate: both scale as $\rmin^{3/2}$ as $\rmin\to 0$. For smaller emission radii $\rmin$, the total flux from the spot decreases while its visibility spreads, essentially undamped, to longer baselines. For example, for the Planck-scale $\rmin$ in Section~\ref{sec:normalization}, the spot visibility is effectively flat across all realistic baselines and forms a constant floor under the photon ring and disk contributions.

\subsection{Physical normalization}
\label{sec:normalization}

The emission mechanism on the surface of last scattering $r = \rmin$ is rather unclear and can probably be determined only in a full theory of quantum gravity. We therefore assume that (i) the emission at $r = \rmin$ follows a Planck law~$B_\nu(T)$ with temperature $T$ in the comoving frame, and (ii) the radius $\rmin$ and the temperature $T$ are free parameters. We then compute the WH spot flux across the $(\rmin, T)$ plane, working in the EHT observing band $\nu_{\mathrm{obs}} = 230\,\mathrm{GHz}$. The natural comparison scale is set by the two EHT sources. At the low accretion rates inferred for M87$^\star$ and Sgr~A$^\star$, the mm emission of the inner accretion flow is synchrotron radiation from hot plasma on horizon scales~\cite{Yuan:2014gma,EventHorizonTelescope:2019pcy} with the observed 230\,GHz compact flux densities $F_{\nu}^{\mathrm{disk}} \approx 0.5\,$Jy for M87$^\star$~\cite{EventHorizonTelescope:2019ths} and $F_{\nu}^{\mathrm{disk}} \approx 2.4\,$Jy for Sgr~A$^\star$~\cite{EventHorizonTelescope:2022ago}.

To relate the unit-emissivity calculations of Section~\ref{sec:signatures} to physical fluxes, recall that the zero-baseline visibility is the integral of the intensity over the image plane (Eq.~\eqref{eq:vis} at $\mathbf{q} = 0$). In other words, $V(q=0)$ gives the total flux density of the image, up to the factor $(M/r_{0})^{2}$ that converts the image-plane area in units of $M^{2}$ into the solid angle subtended on the sky. This means that the curves in Figure~\ref{fig:vis_per_feature} should simply be moved vertically to reflect the correct physical fluxes at zero baseline. 

Now, we can obtain the observed spectrum of the WH spot by using the Lorentz-invariance identity $g^{3} B_{\nu/g}(T) = B_{\nu}(g T)$. Each annulus of impact parameter $b$ radiates as a black body at the locally shifted temperature $g_{\mathrm{spot}}(b)\,T$, so that the total observed flux density is
\begin{equation}\label{eq:F_spot_integral}
    F_{\nu}^{\mathrm{spot}}(\rmin, T) = 2\pi\,\left(\frac{M}{r_{0}}\right)^{2}\!\int_{0}^{\bcrit}\!B_{\nu}\bigl(g_{\mathrm{spot}}(b)\,T\bigr)\,b\,\dd b\,,
\end{equation}
where $T$ is measured in energy units ($k_{B} = 1$), and $\rmin$, $b$, and $\bcrit = 3\sqrt{3}$ are in units of $M$ as before. The integral admits two asymptotic limits, separated by an effective scale
\begin{equation}
\beff = \frac{\rmin T}{h\nu}\,, 
\end{equation}
where $h$ is the Planck constant. The flux density depends on whether the WH spot ($b < \bcrit$) lies fully within that scale or not:
\begin{alignat}{2}
    \label{eq:F_spot_RJ}
    F_{\nu}^{\mathrm{spot}} &\simeq 4\pi\,T\,\nu^{2}\,\rmin\,\bcrit\,\left(\frac{M}{r_{0}}\right)^{2}
        &\qquad &\text{for} \quad \beff \gg \bcrit\,, \\
    \label{eq:F_spot_Wien}
    F_{\nu}^{\mathrm{spot}} &\simeq \frac{2\pi^{3}\,T^{2}\,\nu}{3h}\,\rmin^{2}\,\left(\frac{M}{r_{0}}\right)^{2}
        &\qquad &\text{for} \quad \beff \ll \bcrit\,.
\end{alignat}
In the first case, every annulus of the WH spot radiates in the Rayleigh--Jeans regime, whereas in the second case, the transition from Rayleigh--Jeans to Wien occurs inside the shadow, and the flux is dominated by the central region $b \lesssim \beff$. In both regimes the iso-flux contours follow $\rmin^{p}T^{p} = \text{const}$ with $p \in \{1, 2\}$, giving the same slope $-1$ in the $(\log\rmin, \log T)$ plane. For the full derivation see Appendix~\ref{app:spot_regimes}.

\paragraph*{Planck-star example.} As a model for the values of~$\rmin$ and~$T$, consider the Planck-star scenario~\cite{Rovelli:2014cta,Haggard:2014rza,Barrau:2014hda} (see also~\cite{Markov:1982,Markov:1984ii,Frolov:1998wf,Lukash:2013ts}). The emission radius is set by demanding that the Kretschmann invariant $K(\rmin) = 48/\rmin^{6}$ (with $\rmin$ in units of $M$) has a Planckian value. Its physical units are $[\text{length}^{-4}]$, and we can thus estimate that
\begin{equation}\label{eq:rmin}
    K(\rmin)\sim (\ell_{P}/M)^{-4} \quad \Rightarrow \quad \rmin = 48^{1/6}\,\left(\frac{\ell_{P}}{M}\right)^{\!2/3} \ll 1\,.
\end{equation}
Note that for macroscopic (astrophysical) masses the physical radius $\rmin M$ is well above the actual Planck length $\ell_P\sim 10^{-33}$~cm (e.g., $\rmin M \approx 1.4\times10^{-18}$~cm for $M=10^6M_\odot$). If we also set the emission temperature to its Planck value $T_{P} = (\hbar c^{5}/G)^{1/2} \approx 1.22\times10^{28}\,$eV, Eqs.~\eqref{eq:F_spot_RJ} and~\eqref{eq:F_spot_Wien} read:
\begin{align}
    F_{\nu}^{\mathrm{spot, RJ}} &\simeq 12\sqrt{3}\,\pi\,48^{1/6}\,T_{P}\,\left(\frac{\ell_{P}}{M}\right)^{\!2/3}\,\nu^{2}\,\left(\frac{M}{r_{0}}\right)^{\!2}\,, \\
    F_{\nu}^{\mathrm{spot, Wien}} &\simeq \frac{2\pi^{3}}{3}\,48^{1/3}\,\frac{T_{P}^{2}}{h}\,\left(\frac{\ell_{P}}{M}\right)^{\!4/3}\,\nu\,\left(\frac{M}{r_{0}}\right)^{\!2}\,.
\end{align}
That is, in the Planck-star regime the spot flux is entirely determined by the mass of the eternal black hole and the geometric size of its shadow. At a given observing frequency~$\nu_{\mathrm{obs}}$, the crossover from the Rayleigh--Jeans to Wien formula occurs at the mass scale
\begin{equation}\label{eq:Mstar}
    M_{\star} \approx 10^{8}\,\left(\frac{230\,\mathrm{GHz}}{\nu_{\mathrm{obs}}}\right)^{\!3/2}\,M_{\odot}\,.
\end{equation}
The EHT sources are on the opposite sides of this scale: M87$^\star$ ($M \approx 6.5 \times 10^{9}\,M_{\odot}$) lies in the Wien regime, and Sgr~A$^\star$ ($M \approx 4 \times 10^{6}\,M_{\odot}$) and lower-mass sources (including the intermediate-mass range) are in the Rayleigh--Jeans regime. For the Planck-scale $\rmin$ of Eq.~\eqref{eq:rmin}, the center of the spot is blueshifted by the factor $g_{\mathrm{spot}}(0) \approx \sqrt{2/\rmin} \sim (M/\ell_{P})^{1/3}$. At the same time, the bright core of the spot shrinks to the scale $z = \rmin\sqrt{A} \approx \rmin^{3/2}/\sqrt{2}$ (Section~\ref{sec:signatures}). On the sky, such a spot would appear as a point source with its visibility effectively flat at all realistic baselines. 

\begin{figure}[htbp]
\centering
\includegraphics[width=\textwidth]{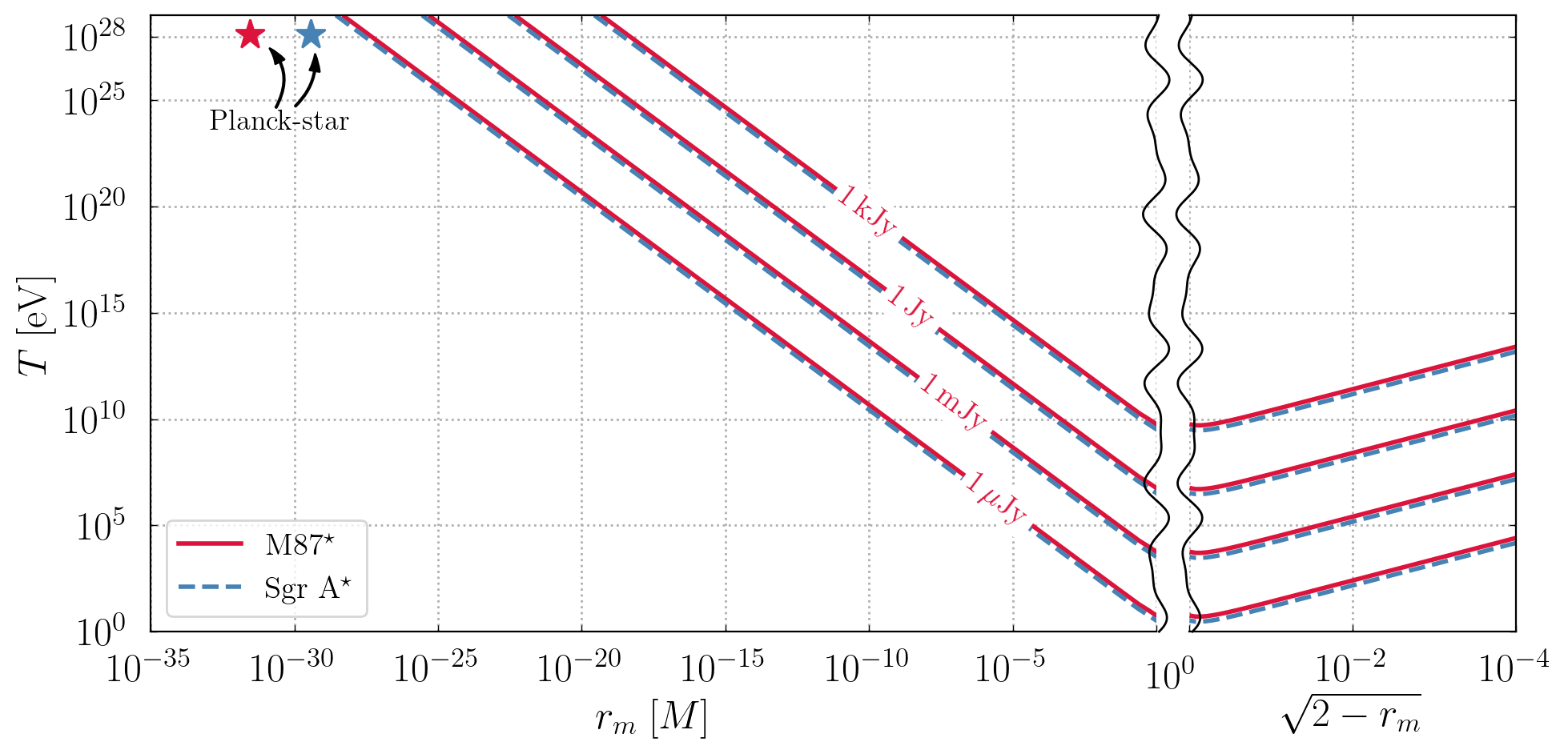}
\caption{Iso-flux contours of the WH spot at 230\,GHz in the $(\rmin, T)$ plane for M87$^\star$ (solid red) and Sgr~A$^\star$ (dashed blue), assuming a black-body emitter at $r = \rmin$ with rest-frame temperature $T$. Contours are spaced by three decades (dex) in flux, from 1\,$\mu$Jy to 1\,kJy. The stars mark the Planck-star regime, Eq.~\eqref{eq:rmin}, for each source. The right panel shows the contours near the horizon, $\rmin\to 2$, where Eqs.~\eqref{eq:F_spot_RJ} and~\eqref{eq:F_spot_Wien} break down, and the exact integral~\eqref{eq:app_F_exact} must be used.}
\label{fig:constraints}
\end{figure}

Figure~\ref{fig:constraints} shows a few iso-flux contours in the $(\rmin,T)$ plane. Once a source's mass and distance are fixed, the respective contours can be obtained from either Eq.~\eqref{eq:F_spot_RJ} or~\eqref{eq:F_spot_Wien}. For illustration we use the same EHT sources, M87$^\star$ and Sgr~A$^\star$, whose ratios $(M/r_{0})^{2}$ are similar ($\approx 15$ and $25\,\mu\mathrm{as}^{2}$, respectively). The lines of constant flux are spaced by three orders of magnitude and span fluxes from 1\,$\mu$Jy to 1\,kJy. We choose the range to be roughly comparable to the capabilities of current and planned interferometric networks. Namely, the instrument sensitivities at 230\,GHz are as follows: $\sim 1$\,Jy for the EHT compact-source level~\cite{EventHorizonTelescope:2019dse}, $\sim 1$\,mJy for ngEHT~\cite{Doeleman:2023kzg}, and $\sim 10\,\mu$Jy for BHEX~\cite{Johnson:2024ttr,Lupsasca:2024xhq}. Two features of Figure~\ref{fig:constraints} stand out. First, the M87$^\star$ and Sgr~A$^\star$ contours are nearly coincident (offset by a mere $0.24$\,dex), as anticipated from the similarity of their $(M/r_{0})^{2}$ ratios. The resulting constraints on $\rmin$ and $T$ are then essentially the same for the two sources. Second, the stars in the figure mark the Planck-star regime for each source, with $\rmin$ given by Eq.~\eqref{eq:rmin} and $T = T_{P}$. Clearly, even the most optimistic detection threshold (BHEX) is approximately three (five) orders of magnitude higher in $\rmin\,T$ than the Planck-star value for Sgr~A$^\star$ (M87$^\star$), which makes this regime undetectable in the near future.

\section{Discussion}
\label{sec:discussion}

In this paper, we have computed the optical appearance of an eternal Schwarzschild black hole surrounded by a geometrically thin accretion disk. We have also recast the Penrose diagrams of the eternal and collapse spacetimes in an observer-friendly form and visualized the null geodesics that we ray-traced with the Luminet method. Astrophysical black holes result from gravitational collapse, and their spacetime contains no WH region. If, however, an eternal black hole describes an object that exists in nature, it is important to establish its observational signature. In the eternal geometry, past-directed light rays with impact parameters below the critical value, $b < \bcrit=3\sqrt{3}$, are continued inside the past WH region and reach the spacelike surface $r = \rmin$ that replaces the past singularity. If this surface emits radiation, there is a bright spot at the center of the ordinary black hole shadow. The profile of this WH spot is determined by the gravitational frequency shift~\eqref{eq:g_spot_body}, with a blueshifted center for $\rmin \lesssim 1$. The spot is a feature that allows one in principle to discriminate between an eternal black hole and one formed by gravitational collapse.

The interferometric signature of the spot turns out to be remarkably simple. The specific intensity of the spot at the observed frequency is a rational function of the impact parameter, Eq.~\eqref{eq:spot_rational}, and the visibility as a function of the baseline length~$q$ takes the closed analytical form $\langle |V(q)|^{2} \rangle \propto e^{-4\pi q z}$, where the decay scale $z$ is set entirely by the emission radius~$\rmin$. For small $\rmin$, both the zero-baseline amplitude and the decay scale behave as $\rmin^{3/2}$, so that a deeper emitting surface makes the spot fainter but its visibility flatter (as a function of the baseline). This exponential decay and the power-law envelopes of the direct disk image ($\propto q^{-3/2}$) and of the photon ring ($\propto q^{-1/2}$) in Figure~\ref{fig:vis_per_feature} are consistent with a general property of the Fourier transform, namely that its decay at long baselines is governed by the smoothness of the image. The disk and the ring contain sharp edges, which produce the power laws, whereas the smooth interior profile of the WH spot decays faster than any power. Its truncation at the photon sphere manifests itself through small quasi-periodic corrections at very long baselines. For $\rmin \ll 1$, the spot acts as a point source, and its visibility contributes an approximately constant flux at all baselines.

The absolute normalization of the spot flux depends on the emission mechanism on the surface of last scattering, which is uncertain and probably requires a full theory of quantum gravity. For this reason, we adopted a minimal model of the black-body emission with temperature~$T$. We then computed the observed flux at 230\,GHz across the $(\rmin, T)$ plane (Figure~\ref{fig:constraints}). The observed compact fluxes of M87$^\star$ and Sgr~A$^\star$ constrain the product $\rmin T$ roughly at the MeV level (recall that $\rmin$ is in units of~$M$). A BHEX level of sensitivity of $10\,\mu$Jy would probe it down to tens of eV. We have also considered the Planck-star scenario, in which $\rmin$ is set by a Planckian value of the Kretschmann invariant and the emission temperature is the Planck temperature, $T = T_{P}$. The corresponding values of $\rmin T$ fall five (M87$^\star$) and three (Sgr~A$^\star$) orders of magnitude below even the BHEX threshold. Therefore, at the current and projected level of sensitivity, the Planck-star regime is out of reach for the two EHT sources.

The main limitation of the present analysis is its restriction to the nonrotating (Schwarzschild) case. Astrophysical black holes are likely to carry angular momentum, and the Kerr geometry introduces qualitative changes to both the shadow shape and the photon ring structure~\cite{Gralla:2020srx,Gralla:2020yvo,Cardenas-Avendano:2023dzo}. The maximal analytic extension of the Kerr metric is an infinite chain of asymptotic regions, and its singularity is a timelike ring rather than a spacelike surface~\cite{Boyer:1966qh,Carter:1968rr}. Therefore, if one is to replace the singularity with a surface as we did in the Schwarzschild case, the anisotropic cosmology argument that we used is unlikely to work. As is also known~\cite{Poisson:1990eh,Ori:1991zz,Dafermos:2017dbw}, the inner (Cauchy) horizon is unstable due to the phenomenon of mass inflation that drives the curvature to Planckian values in the vicinity of the horizon, converting it into a weak null singularity. The classical description of the interior is expected to end there, cutting off the infinite chain of the idealized extension. Although the instability develops toward the future of the Cauchy horizon, the presence of a region of Planckian curvature bounding the classical evolution is consistent with Markov's hypothesis of a limiting density of matter, and it is precisely how we model the surface $r = \rmin$ in the Schwarzschild interior. With all these caveats, we expect the main interferometric prediction to be robust. \textit{As long as the emission coming from the past is smooth within the shadow, the visibility of the resulting spot decays faster than any power of the baseline, due to the smoothness property of the Fourier transform discussed above}.

On the observational side, any emission mechanism that fills the shadow interior could in principle mimic the WH spot. Candidates include optically thin emission from hot gas within the photon sphere (as seen in GRMHD simulations of M87$^\star$ and Sgr~A$^\star$~\cite{EventHorizonTelescope:2019pcy}), jet emission projected onto the shadow region, or more exotic possibilities such as traversable wormholes that transmit light from a second asymptotic region~\cite{Bambi:2013nla,Nedkova:2013msa,Ohgami:2015nra,Shaikh:2018kfv}. A key discriminant is the radial profile. The WH spot has a specific smooth dependence on the impact parameter, which produces an exponentially decaying visibility. Diffuse astrophysical foregrounds, by contrast, typically produce power-law visibility profiles. Time variability offers another test. The eternal geometry is static, and the WH spot is expected to be steady, whereas the hot-gas and jet emission varies on horizon-scale timescales. Multi-frequency observations and polarimetric measurements~\cite{Himwich:2020msm} could further help distinguish these scenarios. A comparison of the observational signatures and visibility profiles of the WH spot and these alternatives would be a natural follow-up to this work.

To conclude, the results of this paper carry two messages. First, if eternal black holes exist in nature, their observation would allow us to look inside the past horizon and receive information about the physical processes in the vicinity of the past singularity. Second, this work provides a way to constrain the existence of such objects, as their shadows would not be fully dark. The absence of a central glow in the observed images would then support the standard picture of astrophysical black holes formed by gravitational collapse.

\noindent\textbf{Acknowledgements.}
This work used Bridges-2 at Pittsburgh Supercomputing Center through allocation PHY260067 from the Advanced Cyberinfrastructure Coordination Ecosystem: Services \& Support (ACCESS) program~\cite{Boerner:2023}, which is supported by U.S.\ National Science Foundation grants \#2138259, \#2138286, \#2138307, \#2137603, and \#2138296.

\appendix

\section{Flux integral for the spot}
\label{app:spot_regimes}

It is convenient to cast the integral~\eqref{eq:F_spot_integral} in an exact dimensionless form. In terms of the constants $A$ and $C$ introduced after Eq.~\eqref{eq:spot_rational} and for an observer at infinity ($C=1$), the frequency-shift factor~\eqref{eq:g_spot_body} is converted into the integration variable
\begin{equation}\label{eq:app_x_def}
    x = \frac{h\nu}{g_{\mathrm{spot}}(b)\,T} = \frac{h\nu}{T}\sqrt{A + \frac{b^2}{\rmin^2}}\,,
\end{equation}
such that the Planck function becomes
\begin{equation}\label{eq:app_Bnu}
    B_{\nu}\bigl(g_{\mathrm{spot}}(b)\,T\bigr) = \frac{2h\nu^{3}/c^{2}}{e^{x} - 1}\,,
\end{equation}
and Eq.~\eqref{eq:F_spot_integral} takes the form
\begin{equation}\label{eq:app_F_exact}
    F_{\nu}^{\mathrm{spot}} = \frac{4\pi h\nu^{3}\,\beff^{\,2}}{c^{2}}\left(\frac{M}{r_{0}}\right)^{2}\int_{\xi_{0}}^{\xi}\frac{x\,\dd x}{e^{x} - 1}\,,
\end{equation}
\begin{equation}\label{eq:xi_beff_defined}
    \xi_{0} = \frac{h\nu}{T}\sqrt{A}\,, \quad \xi = \frac{\bcrit}{\beff}\sqrt{A(\rmin/\bcrit)^2 + 1}\,, \qquad \beff\equiv\frac{\rmin T}{h\nu}\,,
\end{equation}
where the limits $\xi_{0}$ and $\xi$ correspond to $b = 0$ and $b = \bcrit$, respectively. It is this exact form that is used to produce Figure~\ref{fig:constraints}, where it mostly affects the curves in the right panel.

For $\rmin \ll 1$ (and in practice already for $\rmin \lesssim 1$ in the Rayleigh--Jeans regime), the frequency-shift factor reduces to $g_{\mathrm{spot}}(b) \approx \rmin/b$. The lower limit is then $\xi_{0} \approx (h\nu/T)\sqrt{\rmin/2} \approx 0$, while the upper limit becomes $\xi \approx \bcrit/\beff$, and
\begin{equation}\label{eq:app_F_dimensionless}
    F_{\nu}^{\mathrm{spot}} \approx \frac{4\pi h\nu^{3}\,\beff^{\,2}}{c^{2}}\,\left(\frac{M}{r_{0}}\right)^{2}\,\mathcal{I}(\xi)\,,
    \qquad
    \mathcal{I}(\xi) \equiv \int_{0}^{\xi}\frac{x\,\dd x}{e^{x} - 1}\,.
\end{equation}
The dimensionless integral $\mathcal{I}(\xi)$ has two limits:
\begin{equation}\label{eq:app_Ical_limits}
    \mathcal{I}(\xi) \to \xi \quad (\xi \ll 1), \qquad
    \mathcal{I}(\xi) \to \mathcal{I}(\infty) \,=\, \frac{\pi^{2}}{6} \quad (\xi \gg 1)\,,
\end{equation}
which correspond to the Rayleigh--Jeans regime (the Planck integrand $\approx 1$ throughout the shadow) and the Wien regime (the integrand is exponentially cut off for $\beff \lesssim b < \bcrit$), respectively.

Using the Rayleigh--Jeans limit in Eq.~\eqref{eq:app_F_dimensionless} as well as Eq.~\eqref{eq:xi_beff_defined}, we obtain
\begin{equation}\label{eq:app_F_RJ}
    F_{\nu}^{\mathrm{spot, RJ}} \approx \frac{4\pi\,T\,\nu^{2}}{c^{2}}\,\rmin\,\bcrit\,\left(\frac{M}{r_{0}}\right)^{2}\,,
\end{equation}
which reproduces Eq.~\eqref{eq:F_spot_RJ} (with $c=1$). In the Wien limit,
\begin{equation}\label{eq:app_F_Wien_exact}
    F_{\nu}^{\mathrm{spot, Wien}} \approx \frac{4\pi h\nu^{3}\,\beff^{\,2}}{c^{2}}\,\left(\frac{M}{r_{0}}\right)^{2}\cdot\frac{\pi^{2}}{6}
    = \frac{2\pi^{3}\,T^{2}\,\rmin^{2}\,\nu}{3 h c^{2}}\,\left(\frac{M}{r_{0}}\right)^{2}\,,
\end{equation}
which reproduces Eq.~\eqref{eq:F_spot_Wien} (with $c = 1$).

\bibliographystyle{iopart-num}
\bibliography{references}

\end{document}